\documentclass[12pt,letterpaper,twoside]{article} 

\usepackage[dvips]{graphicx,color} 
\usepackage{array,hhline,multicol,multirow,dcolumn} 
\usepackage{rotating} 

\usepackage[DIV13]{typearea}
\usepackage{amsmath}
\usepackage{amssymb}
\usepackage{amsfonts}
\usepackage{amscd}
\usepackage{bbm}
\usepackage[footnotesize]{caption2}
\usepackage{cite}
\usepackage{graphicx}
\usepackage{mathrsfs}
\usepackage[center,footnotesize,hang]{subfigure}

\usepackage{ae,aecompl} 

\newcommand{\inst}[2]{\par#1\it #2}

\begin{document}


\newcommand{\ie}{{\sl i.e.}}
\newcommand{\eg}{{\it e.g.}}
\newcommand{\cf}{{\it cf.}}
\newcommand{\etc}{{\it etc.}}
\newcommand{\eq}{Eq.}
\newcommand{\eqs}{Eqs.}
\newcommand{\fig}{Figure}
\newcommand{\Fig}{Figure}
\newcommand{\figs}{Figures}
\newcommand{\Figs}{Figures}
\newcommand{\tab}{Table}
\newcommand{\tabs}{Tables}
\newcommand{\Tab}{Table}
\newcommand{\Tabs}{Tables}
\newcommand{\Ref}{Ref.}
\newcommand{\Refs}{Refs.}
\newcommand{\Chapter}{Chapter}
\newcommand{\Chapters}{Chapters}
\newcommand{\Sec}{Section}
\newcommand{\Secs}{Sections}
\newcommand{\App}{Appendix}
\newcommand{\Apps}{Appendices}

\newcommand{\equ}[1]{\eq~(\ref{#1})}
\newcommand{\figu}[1]{\fig~\ref{#1}}
\newcommand{\tabl}[1]{\Tab~\ref{#1}}

\newcommand{\scen}[1]{\textsf{\textbf{Scenario #1}}}
\newcommand{\scase}[1]{\textsf{\textbf{Special Case #1}}}

\newcommand{\stheta}{\sin^22\theta_{13}}
\newcommand{\deltacp}{\delta_\mathrm{CP}}
\newcommand{\ldm}{\Delta m_{31}^2} 
\newcommand{\sdm}{\Delta m_{21}^2} 

\newcommand{\JHFSK}{\mbox{\sf T2K}}
\newcommand{\minos}{\mbox{\sf MINOS}}
\newcommand{\icarus}{\mbox{\sf ICARUS}}
\newcommand{\opera}{\mbox{\sf OPERA}}
\newcommand{\CHOOZII}{\mbox{\sf D-Chooz}}
\newcommand{\DChooz}{\mbox{\sf Double-Chooz}}
\newcommand{\ReactorII}{\mbox{\sf Reactor-II}}
\newcommand{\miniboone}{\mbox{\sf MiniBooNE}}
\newcommand{\abr}[1]{{\sc\lowercase{#1}}}

\newcommand{\kov}{\ensuremath{\check\mathrm{C}}erenkov }
\newcommand{\hz}{\ensuremath{\,\mathrm{Hz}}}
\newcommand{\khz}{\ensuremath{\,\mathrm{kHz}}}
\newcommand{\mhz}{\ensuremath{\,\mathrm{MHz}}}
\newcommand{\ev}{\ensuremath{\,\mathrm{eV}}}
\newcommand{\kev}{\ensuremath{\,\mathrm{keV}}}
\newcommand{\mev}{\ensuremath{\,\mathrm{MeV}}}
\newcommand{\gev}{\ensuremath{\,\mathrm{GeV}}}
\newcommand{\cmsq}{\ensuremath{\,\mathrm{cm}^2}}
\newcommand{\nm}{\ensuremath{\,\mathrm{nm}}}
\newcommand{\ms}{\ensuremath{\,\mathrm{m}}}
\newcommand{\msec}{\ensuremath{\,\mathrm{ms}}}
\newcommand{\nsec}{\ensuremath{\,\mathrm{ns}}}
\newcommand{\dmsq}{\ensuremath{\Delta m^{2}~}}
\newcommand{\musec}{\ensuremath{\mu} s}
\newcommand{\numu}{\ensuremath{\nu_{\mu}}}
\newcommand{\numubar}{\ensuremath{\bar\nu_{\mu}}}
\newcommand{\nue}{\ensuremath{\nu_e}}
\newcommand{\nuebar}{\ensuremath{\bar\nu_e}}
\newcommand{\nutau}{\ensuremath{\nu_{\tau}}}
\newcommand{\piplusdecay}{\ensuremath{\pi^+ \rightarrow \mu^+ \numu}}
\newcommand{\piminusdecay}{\ensuremath{\pi^- \rightarrow \mu^- \numubar}}
\newcommand{\pimue}{\ensuremath{\pi \rightarrow \mu \rightarrow e}}
\newcommand{\muplusdecay}{\ensuremath{\mu^+ \rightarrow e^+ \nue \numubar}}
\newcommand{\muminusdecay}{\ensuremath{\mu^- \rightarrow e^- \nuebar \numu}}
\newcommand{\sinsqtheta}{\ensuremath{\sin^2 2 \theta~}}
\newcommand{\mutoe}{\ensuremath{\numu \rightarrow \nue}}
\newcommand{\mutotau}{\ensuremath{\numu \rightarrow \nutau}}
\newcommand{\pizero}{\ensuremath{\pi^{0}}}
\newcommand{\pip}{\ensuremath{\pi^{+}}}
\newcommand{\pim}{\ensuremath{\pi^{-}}}
\newcommand{\pipm}{\ensuremath{\pi^{\pm}}}
\newcommand{\Kp}{\ensuremath{K^{+}}}
\newcommand{\Km}{\ensuremath{K^{-}}}
\newcommand{\KL}{\ensuremath{K^{0}_{L}}}
\newcommand{\KS}{\ensuremath{K^{0}_{S}}}
\newcommand{\reaction}{\ensuremath{\nuebar + p \to e^{+} + n\ \ \ ,}}
\newcommand{\evsq}{\ensuremath{\mathrm{eV}^2}}
\newcommand{\flux}{\,\nu/\mathrm{cm}^{2}/p}
\newcommand{\barflux}{\,\bar \nu/\mathrm{cm}^{2}/p}
\newcommand{\ga}{\gamma}
\newcommand{\gtwid}{\mathrel{\raise.3ex\hbox{$>$\kern-.75em\lower1ex\hbox{$\sim$}}}}
\newcommand{\ltwid}{\mathrel{\raise.3ex\hbox{$<$\kern-.75em\lower1ex\hbox{$\sim$}}}}
\newcommand{\onethree}{\ensuremath{\sin^22\theta_{13}\ }}
\newcommand{\twothree}{\ensuremath{\sin^22\theta_{23}\ }}
\newcommand{\nova}{NO$\nu$A}
%
%
\def\deltamsqthreetwo{$\Delta m_{32}^2$}
\def\nue{$\nu_e$}
\def\numutonue{$\nu_\mu\rightarrow\nu_e$}
\def\signdeltamsqthreetwo{sign$(\Delta_{32}^2)$}
\def\sinsqthetatwothree{$\sin^2 \theta_{23}$}
\def\sinsqtwothetaonethree{$\sin^2 2\theta_{13}$}
\def\sinsqtwothetatwothree{$\sin^2 2\theta_{23}$}
\def\thetaonethree{$\theta_{13}$}
\def\thetatwothree{$\theta_{23}$}

\newcommand{\nuxcarbon}{\ensuremath{\nu_x \hspace{0.05in} ^{12}C \rightarrow \nu_x \hspace{0.05in} ^{12}C^{*}(15.11)}}
\newcommand{\numucarbon}{\ensuremath{\nu_{\mu} \hspace{0.05in} ^{12}C \rightarrow \nu_{\mu} \hspace{0.05in} ^{12}C^{*}(15.11)}}

\newcommand{\CenterObject}[1]{\ensuremath{\vcenter{\hbox{#1}}}}
\newcommand{\CenterEps}[2][1]{\ensuremath{\vcenter{\hbox{\includegraphics[scale=#1]{#2.eps}}}}}

\newcommand{\CiteSeeSaw}{
\cite{Minkowski:1977sc,Yanagida:1980,Glashow:1979vf,Gell-Mann:1980vs,Mohapatra:1980ia}}

\newcommand{\eV}{\ensuremath{\,\mathrm{eV}}}
\newcommand{\keV}{\ensuremath{\,\mathrm{keV}}}
\newcommand{\MeV}{\ensuremath{\,\mathrm{MeV}}}
\newcommand{\GeV}{\ensuremath{\,\mathrm{GeV}}}
\newcommand{\TeV}{\ensuremath{\,\mathrm{TeV}}}

\def\D{\mathrm{d}}
\def\I{\mathrm{i}}
\def\SU{\mathrm{SU}}
\def\U{\mathrm{U}}
\def\ChargeC{\mathrm{C}}
\newcommand{\SuperField}[1]{\bbsymbol{#1}}
\newcommand{\NuR}{N_\mathrm{R}}
\newcommand{\RaiseBrace}[1]{\raise1.5pt\hbox{$\displaystyle#1$}}
%
%
\def\mdmatm{\Delta m^2_{32}}
\def\dmatm{$\mdmatm$}
\def\mdmsol{\Delta m^2_{21}}
\def\dmsol{$\mdmsol$}
\def\numunue{$\nu_\mu \rightarrow \nu_e$}
\def\anumunue{$\bar\nu_\mu \rightarrow \bar\nu_e$}
\def\cerenkov{Cherenkov}
\def\linac{LINAC}

\begin{titlepage}

\begin{center}

\vspace*{10mm}

{\Huge Physics at a Fermilab Proton Driver}

\vspace*{10mm}

\today

FERMILAB-FN-0778-AD-E

\vspace*{10mm}

\newcommand{\anl}      {$^1$}
\newcommand{\bonn}     {$^2$}
\newcommand{\bu}       {$^3$}
\newcommand{\bnl}      {$^4$}
\newcommand{\calabria} {$^5$}
\newcommand{\caltech}  {$^6$}
\newcommand{\uci}      {$^7$}
\newcommand{\cern}     {$^8$}
\newcommand{\chicago}  {$^9$}
\newcommand{\columbia} {$^{10}$}
\newcommand{\desy}     {$^{11}$}
\newcommand{\fnal}     {$^{12}$}
\newcommand{\harvard}  {$^{13}$}
\newcommand{\iit}      {$^{14}$}
\newcommand{\illinois} {$^{15}$}
\newcommand{\indiana}  {$^{16}$}
\newcommand{\jlab}     {$^{17}$}
\newcommand{\karlsruhe}{$^{18}$}
\newcommand{\kek}      {$^{19}$}
\newcommand{\kvi}      {$^{20}$}
\newcommand{\lanl}     {$^{21}$}
\newcommand{\lsu}      {$^{22}$}
\newcommand{\madrid}   {$^{23}$}
\newcommand{\tum}      {$^{24}$}
\newcommand{\nwu}      {$^{25}$}
\newcommand{\ornl}     {$^{26}$}
\newcommand{\okstate}  {$^{27}$}
\newcommand{\odu}      {$^{28}$}
\newcommand{\osaka}    {$^{29}$}
\newcommand{\pisa}     {$^{30}$}
\newcommand{\princeton}{$^{31}$}
\newcommand{\ias}      {$^{32}$}
\newcommand{\rutgers}  {$^{33}$}
\newcommand{\slac}     {$^{34}$}
\newcommand{\tech}     {$^{35}$}
\newcommand{\tn}       {$^{36}$}
\newcommand{\trieste}  {$^{37}$}
\newcommand{\triumf}   {$^{38}$}
\newcommand{\virginia} {$^{39}$}
\newcommand{\wisconsin}{$^{40}$}
\newcommand{\yale}     {$^{41}$}

\small
M.G.~Albrow\fnal,              
S.~Antusch\madrid,             
K.S.~Babu\okstate,             
T.~Barnes\tn$^,$\ornl,         
A.O.~Bazarko\princeton,        
R.H.~Bernstein\fnal,           
T.J.~Bowles\lanl,              
S.J.~Brice\fnal,               
A.~Ceccucci\cern,              
F.~Cei\pisa,                   
H.W.K~Cheung\fnal,             
D.C.~Christian\fnal,           
J.I.~Collar\chicago,           
J.~Cooper\fnal,                
P.S.~Cooper\fnal,              
A.~Curioni\yale,               
A.~de~Gouv\^{e}a\nwu,          
F.~DeJongh\fnal,               
P.F.~Derwent\fnal,             
M.V.~Diwan\bnl,                
B.A.~Dobrescu\fnal,            
G.J.~Feldman\harvard,          
D.A.~Finley\fnal,              
B.T.~Fleming\yale,             
S.~Geer\fnal,                  
G.L.~Greene\ornl$^,$\tn,       
Y.~Grossman\tech$^,$\bu$^,$\harvard,   
D.A.~Harris\fnal,              
C.J.~Horowitz\indiana,         
D.W.~Hertzog\illinois,         
P.~Huber\wisconsin,            
J.~Imazato\kek,                
A.~Jansson\fnal,               
K.P.~Jungmann\kvi,             
P.A.~Kasper\fnal,              
J.~Kersten\desy,               
S.H.~Kettell\bnl,              
Y.~Kuno\osaka,                 
M.~Lindner\tum,                
M.~Mandelkern\uci,             
W.J.~Marciano\bnl,             
W.~Melnitchouk\jlab,           
O.~Mena\fnal,                  
D.G.~Michael\caltech,          
J.P.~Miller\bu,                
G.B.~Mills\lanl,               
J.G.~Morf\'{i}n\fnal,          
H.~Nguyen\fnal,                
U.~Nierste\fnal$^,$\karlsruhe, 
T.~Numao\triumf,               
A.~Para\fnal,                  
S.J.~Parke\fnal,               
D.~Pocanic\virginia,           
A.~Psaker\odu$^,$\jlab,        
R.D.~Ransome\rutgers,          
M.~Ratz\bonn,                  
R.E.~Ray\fnal,                 
B.L.~Roberts\bu,               
W.~Roberts\odu,                
M.~Rolinec\tum,                
A.~Sato\osaka,                 
T.~Schwetz\trieste,            
V.~Shiltsev\fnal,              
N.~Solomey\iit,                
T.M.P.~Tait\anl,               
R.~Tayloe\indiana,             
R.~Tschirhart\fnal,            
V.L.~Tumakov\uci,              
R.G.~Van~de~Water\lanl,        
G.~Violini\calabria,           
Y.W.~Wah\chicago,              
M.O.~Wascko\lsu,               
W.~Winter\ias,                 
M.~Yamaga\osaka,               
T.~Yamanaka\osaka,             
G.P.~Zeller\columbia           

\newpage

\inst{\anl}{Argonne National Laboratory, Argonne, Illinois 60439}               
\inst{\bonn}{Physikalisches Institut der Universit\"{a}t Bonn, 53115 Bonn, Germany} 
\inst{\bu}{Boston University, Boston, Massachusetts 02215}                      
\inst{\bnl}{Brookhaven National Laboratory, Upton, New York 11973}              
\inst{\calabria}{Universit\`{a} della Calabria, Cosenza, I-87036 Italy and INFN, Gruppo Collegato di Cosenza, LNF, Italy}         
\inst{\caltech}{California Institute of Technology, Pasadena, California 91125} 
\inst{\uci}{University of California at Irvine, Irvine, California 92697}       
\inst{\cern}{CERN, CH-1211, Gen\`{e}ve 23, Switzerland}                         
\inst{\chicago}{University of Chicago, Chicago, Illinois 60637}                 
\inst{\columbia}{Columbia University, New York, New York 10027}                 
\inst{\desy}{Deutsches Elektronen-Synchrotron (DESY), 22603 Hamburg, Germany}   
\inst{\fnal}{Fermi National Accelerator Laboratory, Batavia, Illinois 60510}    
\inst{\harvard}{Harvard University, Cambridge, Massachusetts 02138}             
\inst{\iit}{Illinois Institute of Technology, Chicago, Illinois 60616}          
\inst{\illinois}{University of Illinois at Urbana-Champaign, Illinois 61801}    
\inst{\indiana}{Indiana University, Bloomington, Indiana 47405}                 
\inst{\jlab}{Thomas Jefferson National Accelerator Facility, Newport News, Virgina 23606}  
\inst{\karlsruhe}{Universit\"at Karlsruhe, 76128 Karlsruhe, Germany} 
\inst{\kek}{IPNS, High Energy Accelerator Research Organization (KEK), Ibaraki 305-0801, Japan} 
\inst{\kvi}{Kernfysisch Versneller Instituut, Zernikelaan 25, NL-9747 AA Groningen, Netherlands} 
\inst{\lanl}{Los Alamos National Laboratory, Los Alamos, New Mexico 87545}      
\inst{\lsu}{Louisiana State University, Baton Rouge, Louisiana 70803}           
\inst{\madrid}{Universidad Aut\'onoma de Madrid, 28049 Madrid, Spain}           
\inst{\tum}{Technische Universit\"{a}t M\"{u}nchen, 85748 Garching, Germany}    
\inst{\nwu}{Northwestern University, Evanston, Illinois 60208}                  
\inst{\ornl}{Oak Ridge National Laboratory, Oak Ridge, Tennessee 37831}         
\inst{\okstate}{Oklahoma State University, Stillwater, Oklahoma 74078}          
\inst{\odu}{Old Dominion University, Norfolk, Virginia 23529}                   
\inst{\osaka}{Osaka University, Toyonaka, Osaka 560-0043, Japan}                
\inst{\pisa}{Dipartimento di Fisica dell'Universit\`{a} di Pisa and INFN, 56010 Pisa, Italy} 
\inst{\princeton}{Princeton University, Princeton, New Jersey 08544}            
\inst{\ias}{Institute for Advanced Study, Princeton New Jersey 08540}           
\inst{\rutgers}{Rutgers University, Piscataway, New Jersey 08855}               
\inst{\tech}{Technion-Israel Institute of Technology, Technion City, 32000 Haifa, Israel} 
\inst{\tn}{University of Tennessee, Knoxville, Tennessee 37996}                 
\inst{\trieste}{Scuola Internazionale Superiore di Studi Avanzati, I-34014 Trieste, Italy} 
\inst{\triumf}{TRIUMF, Vancouver, British Columbia V6T 2A3 Canada}              
\inst{\virginia}{University of Virginia, Charlottesville, Virginia 22901}       
\inst{\wisconsin}{University of Wisconsin, Madison, Wisconsin 53706}            
\inst{\yale}{Yale University, New Haven, Connecticut 06520}                     

\end{center}

\end{titlepage}

\clearpage

\section*{Executive Summary}

In the last few years there has been interest in a new generation 
of high intensity multi-GeV proton accelerators. 
At Fermilab, two possible proton driver schemes have been proposed to 
enable the Main Injector (MI) beam power to be increased by about a factor of five to 
2~megawatts. The presently favored scheme is based on a new 8~GeV 
superconducting (SC) linac that utilizes, and helps develop, Linear Collider 
technology.

The interest in a new Fermilab Proton Driver is  
motivated by the exciting discoveries that have been made in the 
neutrino sector; namely that neutrinos have mass and that neutrinos 
of one flavor can transform themselves into neutrinos of a different 
flavor as they propagate over macroscopic distances. This is exciting 
because it requires new physics beyond the Standard Model. However, 
we do not yet have a complete knowledge of neutrino masses and mixing. 
Understanding these neutrino properties is important 
because neutrinos are the most common matter particles in the universe. In number, 
they exceed the constituents of ordinary matter (electrons, protons, neutrons) 
by a factor of ten billion. They probably account for at least as much energy in the 
universe as all the stars combined and, depending on their exact masses, might 
also account for a few percent of the so-called ``dark matter''. In addition, 
neutrinos are important in stellar processes. There are 
70 billion per second 
streaming through each square centimeter of the Earth from the Sun.  
Neutrinos govern the dynamics of supernovae, and hence the production of heavy 
elements in the universe. Furthermore, if there is CP violation in the neutrino 
sector, the physics of neutrinos in the early universe might ultimately be 
responsible for baryogenesis.
{\it If we are to understand ``why we are here'' and the basic nature of the 
universe in which we live, we must understand the basic properties of the 
neutrino}.

To identify the best ways to address the most important open neutrino 
questions, and to determine an effective, fruitful U.S. role within a 
global experimental neutrino program, the American Physical Society's 
Divisions of Nuclear Physics and Particles and Fields, together with the 
Divisions of Astrophysics and the Physics of Beams, have recently conducted 
a ``Study on the Physics of Neutrinos''. This study recommended  
{\it ``... as a high priority, a comprehensive U.S. program to 
complete our understanding of neutrino mixing, to determine the character 
of the neutrino mass spectrum, and to search for CP violation among 
neutrinos'' }, and identified, as a key ingredient of the 
future program,  
{\it ``A proton driver in the megawatt class or above and neutrino superbeam 
with an appropriate very large detector capable of observing CP violation 
and measuring the neutrino mass-squared differences and mixing parameters 
with high precision.'' } The proposed Fermilab Proton Driver would, 
together with a suitable new detector,  
fulfill this need by providing a 2~megawatt proton beam at Main Injector 
(MI) energies for the future ``Neutrinos at the Main Injector'' (NuMI) program.

The NuMI beam is unique. It is the only neutrino beam that has an appropriate 
energy and a sufficiently long baseline to produce, due to matter effects, 
significant changes in the effective oscillation parameters. These matter 
effects can be exploited to determine the pattern of neutrino masses. 
Furthermore, when combined with measurements from the much-shorter-baseline 
T2K experiment being built in Japan,
an appropriate NuMI-based experiment could exploit matter effects to achieve 
a greatly enhanced sensitivity to CP violation in the neutrino sector. 

To obtain sufficient statistical sensitivity to determine the pattern of 
neutrino masses and search for CP violation over a large region 
of parameter-space will require a new detector with a fiducial mass of tens
of kilotons and a neutrino beam with the highest practical intensity. 
Hence, the primary motivation for the new Fermilab Proton Driver is to 
enable an increase in the MI beam power to the maximum that is considered 
practical. The achievable sensitivity to the pattern of 
neutrino masses, and to CP violation, will depend on the unknown neutrino 
mixing angle $\theta_{13}$. Experiments using the NuMI beam in the Fermilab 
Proton Driver era would be able to search for a finite $\theta_{13}$ with 
a sensitivity well beyond that achievable with the present NuMI beam, the 
T2K beam, or at future reactor experiments.

In the presently favored 8~GeV SC linac proton driver scheme the MI fill 
time is very short (<1~ms), which means that the MI can deliver 
2~megawatts of beam at any energy from 40 to 120~GeV, and improvements to 
the MI ramp 
time can further increase the beam power. The short fill time also means that 
the majority of the 8~GeV cycles will not be used by the MI. Hence the 
SC linac could support a second high-intensity physics program using the 
primary beam at 8~GeV with an initial beam power of 0.5~megawatts, 
upgradeable to 2~megawatts (a factor of 64 increase of the present 8 GeV Booster beam).
Both the primary proton beams (MI and 8~GeV) 
could be used to create neutrino beams. Both these beams are needed for 
an extensive program of neutrino scattering measurements. These 
measurements are not only of interest in their own right, but are also 
needed to reduce the systematic uncertainties on the neutrino oscillation 
measurements which arise from our limited knowledge of the 
relevant neutrino cross sections.

Although neutrino oscillations provide the primary motivation for interest in 
the Fermilab Proton Driver, the participation in recent 
proton driver physics workshops has been broader than the neutrino physics 
community. Note that intense muon, pion, kaon, neutron, and antiproton 
beams at the Fermilab Proton Driver would offer great flexibility for 
the future program, and could support a diverse program of experiments of 
interest to particle physicists, nuclear physicists, and 
nuclear-astrophysicists.
In particular, as the Large Hadron Collider (LHC) 
and International Linear Collider (ILC) begin to 
probe the energy frontier, a new round of precision flavor physics experiments 
would provide information that is complementary to the LHC and ILC data by indirectly 
probing high mass scales through radiative corrections. This would help to 
elucidate the nature of any new physics that is discovered at the energy frontier. 
Examples of specific experiments of this type that could be supported at the 
Fermilab Proton Driver include 
(i) at the MI: $K^+ \rightarrow \pi^+ \nu \overline{\nu}$, 
$K_L \rightarrow \pi^0 \nu \overline{\nu}$, and 
$K_L \rightarrow \pi^0 e^+ e^-$, and 
(ii) using the 8~GeV primary beam to produce an intense low energy muon source: 
muon $(g-2)$ measurements and searches for a muon electric dipole moment, 
$\mu \rightarrow e \gamma$, and $\mu \rightarrow e$ conversion. 
Should no new physics be discovered at the LHC and/or ILC then, for the 
foreseeable future, precision muon, pion, kaon, and neutron measurements 
at a high-intensity proton source may provide the only practical way to 
probe physics at higher mass scales.

\clearpage

The main conclusions presented in this report are:

\begin{enumerate}
\item Independent of the value of the unknown mixing angle $\theta_{13}$ 
an initial Fermilab Proton Driver long-baseline neutrino experiment 
will make a critical contribution to the global oscillation program.
\item  If $\theta_{13}$ is very small the initial Fermilab Proton Driver 
experiment will provide the most stringent limit on 
$\theta_{13}$ and prepare the way for a neutrino factory. The 
expected $\theta_{13}$ sensitivity exceeds that expected for reactor-based 
experiments, or any other accelerator-based experiments.  
\item If $\theta_{13}$ is sufficiently large the initial Fermilab Proton 
Driver experiment will precisely measure its value, 
perhaps determine the mass hierarchy, and prepare the way for a sensitive 
search for CP violation. The value of $\theta_{13}$ will guide the 
further evolution of the Proton Driver neutrino program. 
\item The Fermilab Proton Driver neutrino experiments will also make 
precision measurements of the other oscillation parameters, and conduct 
an extensive set of neutrino scattering measurements, some of which are 
important for the oscillation program. Note that the neutrino scattering 
measurements require the highest achievable intensities at both MI energies and 
at 8~GeV.
\item The Fermilab Proton Driver could also support a broad range of other 
experiments of interest to particle physicists, nuclear physicists, and 
nuclear astrophysicists. These experiments could exploit antiproton- 
and kaon-beams from the MI, or muon-, pion-, or neutron-beams from 
the 8~GeV linac. These ``low energy'' experiments would provide 
sensitivity to new physics at high mass scales which would be complementary to 
measurements at the LHC and beyond.
\end{enumerate}

\clearpage

\tableofcontents

\cleardoublepage

\section{Introduction}
\label{ch:Introduction}

In the last few years there has been interest in a new generation 
of high intensity multi-GeV proton accelerators capable of delivering 
a beam of one or a few megawatts. The interest in these high-intensity 
accelerators is driven by the exciting discoveries that have been made in the 
neutrino sector; namely that neutrinos have mass and that neutrinos of one 
flavor can transform themselves into neutrinos of a different flavor as they 
propagate over macroscopic distances. This requires new physics beyond the 
Standard Model (SM). To identify the most important open neutrino physics questions, 
evaluate the physics reach of various proposed ways of answering the questions, 
and to determine an effective, fruitful U.S. role within a global experimental 
neutrino program, the American Physical Society's Divisions of Nuclear Physics 
and Particles and Fields, together with the Divisions of Astrophysics and the 
Physics of Beams, have recently conducted a ``Study on the Physics of 
Neutrinos''. The resulting APS report~\cite{the-neutrino-matrix} recommended  
{\it ``... as a high priority, a comprehensive U.S. program to 
complete our understanding of neutrino mixing, to determine the character 
of the neutrino mass spectrum, and to search for CP violation among 
neutrinos.'' } The APS study identified, as a key ingredient of the 
future program,  
{\it ``A proton driver in the megawatt class or above and neutrino superbeam 
with an appropriate very large detector capable of observing CP violation 
and measuring the neutrino mass-squared differences and mixing parameters 
with high precision.'' } A Fermilab Proton Driver would, 
together with a suitable new detector,  
fulfill this need by providing a 2~megawatt proton beam at Main Injector 
(MI) energies for the future ``Neutrinos at the MI'' (NuMI) program.

Fermilab hosts the U.S. accelerator-based neutrino program and, with the 
recently completed NuMI beamline, is operating the longest-baseline neutrino 
beam in the world. The NuMI beam will, for the foreseeable future, provide the 
only accelerator-based neutrino baseline that is long enough for matter effects 
to significantly change the effective neutrino oscillation parameters. These 
matter effects can be exploited to answer one of the key questions in neutrino 
physics, namely: Which of the two presently viable patterns of neutrino mass 
is the correct one ?  Furthermore, when combined with measurements from the much 
shorter-baseline T2K experiment being built in Japan,
an appropriate NuMI-based experimental program could exploit matter effects and achieve 
a greatly enhanced sensitivity to CP violation in the neutrino sector. 

To obtain sufficient statistical sensitivity to determine the pattern of 
neutrino masses and search for CP violation over a large region of parameter
-space will require a new detector with a fiducial mass of a few tens of kilotons, 
and a neutrino beam with the highest practical intensity. Hence, the primary 
motivation for the new Fermilab Proton Driver is to provide an increase in the MI 
beam power to the maximum that is considered practical, namely 2~megawatts. 
The achievable sensitivity to the pattern of neutrino masses, and 
to CP violation, will depend on the unknown neutrino mixing angle $\theta_{13}$. 
Experiments using the NuMI beam in the Fermilab Proton Driver era would be able to 
search for a finite $\theta_{13}$ with a sensitivity well beyond that achievable 
with the present NuMI beam, the T2K beam, or at future reactor experiments.
For this reason the APS neutrino study report recommended a new proton driver 
be constructed as early as is practical. In the illustrative road map given in 
the APS report, construction begins in 2008, with operation beginning in 2012. 
In the longer term, should $\theta_{13}$ turn out to be close to or beyond the 
limiting sensitivity of the first round of Fermilab Proton Driver experiments, the 
Fermilab Proton Driver would offer options for further upgrades to the detector and/or 
beamline to yield another stepwise improvement in sensitivity. There would also 
be an option to develop the Fermilab Proton Driver complex to support a neutrino 
factory. 

The preferred Fermilab Proton Driver scheme is based on a new 8~GeV superconducting (SC) linac that 
utilizes, and helps develop, Linear Collider technology. The MI fill time is 
very short ($< 1$~ms), which means that the MI can deliver 2~megawatts of 
beam at any energy from 40 to 120~GeV, and that improvements to the MI ramp 
time can further increase the beam power. The short fill time also means that 
the majority of the 8~GeV cycles will not be used by the MI. Hence the SC linac 
could support a second high-intensity physics program using the primary beam 
at 8~GeV with an initial beam power of 0.5~megawatts, upgradeable to 2~megawatts. 

Although neutrino oscillations provide the primary motivation for interest in 
the Fermilab Proton Driver, the community participating in recent 
proton driver physics workshops has been broader than the neutrino physics 
community. Note that intense muon, pion, kaon, neutron, and antiproton 
beams at the Fermilab Proton Driver would offer great flexibility for 
the future program, and could support a diverse program of experiments of 
interest to particle physicists, nuclear physicists, and 
nuclear-astrophysicists.
In particular, as the Large Hadron Collider (LHC) 
and International Linear Collider (ILC) begin to 
probe the energy frontier, a new round of precision flavor physics experiments 
would provide information that is complementary to the LHC and ILC data by indirectly 
probing high mass scales through radiative corrections. This would help to 
elucidate the nature of any new physics that is discovered at the energy frontier. 
Examples of specific experiments of this type that could be supported at the 
Fermilab Proton Driver include 
(i) at the MI: $K^+ \rightarrow \pi^+ \nu \overline{\nu}$, 
$K_L \rightarrow \pi^0 \nu \overline{\nu}$, and 
$K_L \rightarrow \pi^0 e^+ e^-$, and 
(ii) using the 8~GeV primary beam to produce an intense low energy muon source: 
muon $(g-2)$ measurements and searches for a muon electric dipole moment, 
$\mu \rightarrow e \gamma$, and $\mu \rightarrow e$ conversion. 
Should no new physics be discovered at the LHC and/or ILC then, for the 
foreseeable future, precision muon, pion, kaon, and neutron measurements 
at a high-intensity proton source may provide the only practical way to 
probe physics at higher mass scales.

This document summarizes the physics opportunities that would be provided by a new 
proton driver at Fermilab. In particular, the physics that could be done with 
a 2~megawatt MI beam, and the physics that could be done with a 0.5 - 2~megawatt 
8~GeV beam. Sections 2 and 3 describe respectively the potential neutrino 
oscillation and neutrino scattering physics programs. Section~4 describes the 
broader physics program using muon-, pion-, and neutron-beams produced with a high 
intensity primary proton beam at 8~GeV, and using 
kaon- and antiproton-beams produced with the MI primary proton beam.
 An overview of the complete proton driver 
program is given in Section~5, and a summary in Section~6.

\cleardoublepage

\section{Neutrino Oscillations}
\label{ch:Neutrinos}

Neutrinos are the most common matter particles in the universe. In number, 
they exceed the constituents of ordinary matter (electrons, protons, neutrons) 
by a factor of ten billion. They probably account for at least as much energy in the 
universe as all the stars combined and, depending on their exact masses, might 
also account for a few percent of the so-called ``dark matter''. In addition, 
neutrinos are important in stellar processes. There are about 
$7 \times 10^{10}$ cm$^{-2}$ sec$^{-1}$ streaming through the Earth from the Sun.  
Neutrinos govern the dynamics of supernovae, and hence the production of heavy 
elements in the universe. Furthermore, if there is CP violation in the neutrino 
sector, the physics of neutrinos in the early universe might ultimately be 
responsible for baryogenesis.
{\it If we are to understand ``why we are here'' and the basic nature of the 
universe in which we live, we must understand the basic properties of the 
neutrino}.

In the last few years solar, atmospheric, and reactor neutrino experiments have 
revolutionized our understanding of the nature of neutrinos. We now know that 
neutrinos produced in a given flavor eigenstate can transform themselves into 
neutrinos of a different flavor as they propagate over macroscopic 
distances. This means that, like quarks, neutrinos have a non-zero mass, the flavor 
eigenstates are different from the mass eigenstates, and hence neutrinos mix. 
However, we have incomplete knowledge of the properties of neutrinos since 
{\it we do not know the spectrum of neutrino masses, and we have only partial 
knowledge of the mixing among the three known 
neutrino flavor eigenstates}.  Furthermore, it is possible that the simplest 
three-flavor mixing scheme is not the whole story, and that a complete 
understanding of neutrino properties will require a more complicated framework. 
In addition to determining the parameters that describe the neutrino sector, 
the three-flavor mixing framework must also be tested.

The SM cannot accommodate non-zero neutrino mass terms without some 
modification. We must either introduce right-handed neutrinos (to generate 
Dirac mass terms) or allow neutrinos to be their own antiparticle (violating 
lepton number conservation, and allowing Majorana mass terms). Hence 
{\it the physics of neutrino masses is physics beyond the Standard Model}.
Although we do not know the neutrino mass spectrum, we do know that the masses, 
and the associated mass-splittings, are tiny compared to the masses of any other 
fundamental fermion. This suggests that the physics responsible for neutrino mass 
will include new components radically different from those of the SM. 
Furthermore, although we do not have complete knowledge of the mixing between different neutrino flavors, we do know that it is qualitatively very different from the corresponding mixing between different quark flavors. The observed difference necessarily constrains our ideas 
about the underlying relationship between quarks and leptons, and hence models 
of quark and lepton unification in general, and Grand Unified Theories (GUTs) in 
particular. Note that in neutrino mass models the seesaw mechanism \cite{Minkowski:1977sc,Yanagida:1980,Glashow:1979vf,Gell-Mann:1980vs,Mohapatra:1980ia} provides a 
quantitative explanation for the observed small neutrino masses, which arise as 
a consequence of the existence of right-handed neutral leptons at the GUT-scale. 
Over the last few years, as our knowledge of the neutrino oscillation parameters 
has improved, a previous generation of neutrino mass models has already been ruled 
out, and a new set of models has emerged specifically designed to accommodate the 
neutrino parameters. Further improvement in our knowledge of the oscillation 
parameters will necessarily reject many of these models, and presumably encourage 
the emergence of new ideas. Hence {\it neutrino physics is experimentally driven, 
and the experiments are already directing our ideas about what lies beyond the 
Standard Model}.

Our desire to understand both the universe in which we live and physics beyond 
the SM provides a compelling case for an experimental program that 
can elucidate the neutrino mass spectrum, measure neutrino mixing, and test the 
three-flavor mixing framework. It seems likely that complete knowledge of the 
neutrino mass splittings and mixing parameters is accessible to 
accelerator-based neutrino oscillation experiments. In the following we first 
introduce the three-flavor mixing framework and identify the critical 
measurements that need to be made in the future oscillation physics program. 
The sensitivity of the Fermilab program based on a new Proton Driver is 
then considered in the context of the global experimental program.

\subsection{Oscillation Framework and Measurements}
\label{sec:osc_framework}

There are three known neutrino flavor eigenstates 
$\nu_\alpha = (\nu_e, ~\nu_\mu, ~\nu_\tau)$. Since transitions have been 
observed between the flavor eigenstates we now know that neutrinos have 
non-zero masses, and that there is mixing between the flavor eigenstates.
The mass eigenstates $\nu_i = (\nu_1, ~\nu_2, ~\nu_3)$ with masses 
$m_i = (m_1, ~m_2, ~m_3)$ are related to the flavor eigenstates by 
a $3 \times 3$ unitary mixing matrix $U^\nu$, 
\begin{equation}
  |\nu_\alpha\rangle = \sum_i ( U^\nu_{\alpha i} )^* |\nu_i\rangle
\label{mix}
\end{equation}
Four numbers are needed to specify all of the matrix elements, namely three mixing 
angles ($\theta_{12}, \theta_{23}, \theta_{13}$) and one complex 
phase ($\delta$). In terms of these parameters
\begin{equation}
\hspace*{-0.5cm}
U^\nu =
\left( \begin{array}{ccc}
  c_{13} c_{12}       & c_{13} s_{12}  & s_{13} e^{-i\delta} \\
- c_{23} s_{12} - s_{13} s_{23} c_{12} e^{i\delta}
& c_{23} c_{12} - s_{13} s_{23} s_{12} e^{i\delta}
& c_{13} s_{23} \\
    s_{23} s_{12} - s_{13} c_{23} c_{12} e^{i\delta}
& - s_{23} c_{12} - s_{13} c_{23} s_{12} e^{i\delta}
& c_{13} c_{23}
\end{array} \right) \,
\label{mns}
\end{equation}
where $c_{jk} \equiv \cos\theta_{jk}$ and $s_{jk} \equiv \sin\theta_{jk}$. 
Neutrino oscillation measurements have already provided some knowledge 
of $U^\nu$, which is approximately given by:
\begin{equation}
\hspace*{-0.5cm}
U^\nu =
\left( \begin{array}{ccc}
  0.8  & 0.5  & ? \\
  0.4  & 0.6  & 0.7 \\
  0.4  & 0.6  & 0.7 \\
\end{array} \right) \,
\label{mnsnumbers}
\end{equation}
We have limited knowledge of the (1,3)-element of the mixing matrix. This 
matrix element is parametrized by $s_{13} e^{-i\delta}$. We have only an upper 
limit on $\theta_{13}$ and no knowledge of $\delta$. 
Note that $\theta_{13}$ and $\delta$ are particularly important because if 
$\theta_{13}$ and $\sin \delta$ are non-zero there will be CP violation in 
the neutrino sector.

\begin{figure}[t]
\begin{center}
\includegraphics[width=\textwidth]{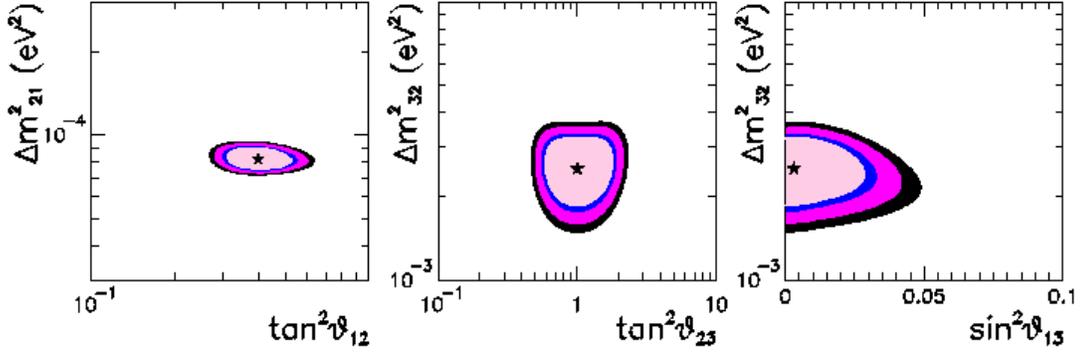}
\end{center}
\caption{\label{parsfig} Current experimental constraints on the three mixing angles $\theta_{12}$, $\theta_{23}$, and $\theta_{13}$ and their dependence on the two known mass-squared differences $\Delta m_{12}^2$ and $\Delta m_{23}^2$. The star indicates the most likely solution. Figure taken from \cite{the-neutrino-matrix}.}
\end{figure}

\begin{figure}[t]
\begin{center}
\includegraphics[width=0.9\textwidth]{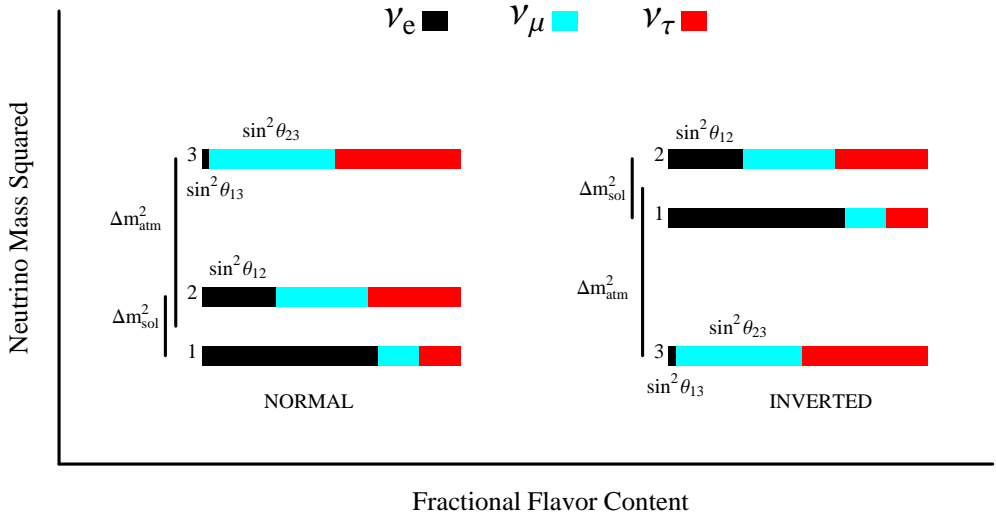}
\end{center}
\caption{\label{hierfig}
The two possible arrangements of the masses of the three known neutrinos, 
based on neutrino oscillation measurements. The spectrum on the left 
corresponds to the {\it Normal Hierarchy} and has $\Delta m^2_{32} > 0$. 
The spectrum on the right corresponds to the {\it Inverted Hierarchy} and 
has $\Delta m^2_{32} < 0$.
The $\nu_e$ fraction of each mass eigenstate is indicated by the black solid
region. The $\nu_\mu$ and $\nu_\tau$ fractions are indicated by
the blue (red) regions respectively.
The $\nu_e$ fraction in the
mass eigenstate labeled ``3'' has been set to the CHOOZ bound. Figure taken from \cite{olga}.}
\end{figure}
Neutrino oscillations are driven by the splittings between the neutrino 
mass eigenstates. 
It is useful to define the differences between the squares of the masses of 
the mass eigenstates $\Delta m^2_{ij} \equiv m^2_i-m^2_j$. 
The probability that a neutrino of energy $E$ and initial flavor $\alpha$ will 
``oscillate'' into a neutrino of flavor $\beta$ is given by  
$P_{\alpha \beta} \equiv P(\nu_\alpha \rightarrow \nu_\beta) =
\left| \langle \nu_\beta | \exp( - i \mathcal{H} t ) | \nu_\alpha \rangle \right|^2$,
which in vacuum is given by 
\begin{equation}
P_{\alpha \beta}  =  \left| \sum\limits_{j=1}^3 U_{\alpha j}^* U_{\beta j} \, \exp(- i E_j t) \right|^2  
=   \sum\limits_{j=1}^3 \, \sum\limits_{k=1}^3 U_{\alpha j} U_{\alpha k}^* U_{\beta j}^* U_{\beta k} \, 
\exp \left( - i \frac{\Delta m_{kj}^2}{2 E} t \right)
\label{equ:oscgeneral}
\end{equation}
If neutrinos of energy $E$ travel a distance $L$ then a measure of the propagation time 
$t$ is given by $L/E$. Non-zero $\Delta m^2_{ij}$ will result in neutrino flavor 
oscillations that have maxima at given values of $L/E$, and oscillation 
amplitudes that are determined by the matrix elements $U^\nu_{\alpha i}$, and 
hence by $\theta_{12}, \theta_{23}, \theta_{13}$, and $\delta$. 

Our present knowledge of the neutrino mass splittings and mixing matrix, has been 
obtained from atmospheric \cite{Fukuda:2000np,Fukuda:1998mi}, solar \cite{Lande:2003ex,Abdurashitov:1999bv,Hampel:1998xg,Ahmad:2001an,Ahmed:2003kj,Fukuda:2002pe}, reactor \cite{Apollonio:1999ae,Boehm:2001ik,Eguchi:2002dm}, and 
accelerator-based \cite{Aliu:2004sq} neutrino experiments, and is summarized in Fig.~\ref{parsfig}. The 
solar-neutrino experiments and the reactor experiment KamLAND probe values of 
$L/E$ that are sensitive to $\Delta m^2_{21}$, and the mixing angle $\theta_{12}$. 
Our knowledge of these parameters is shown in the left panel of Fig.~\ref{parsfig}.
The atmospheric-neutrino experiments and the accelerator based experiment K2K probe 
values of $L/E$ that are sensitive to $\Delta m^2_{32}$, and the mixing angle 
$\theta_{23}$. Our knowledge of these parameters is shown in the central panel of Fig.~\ref{parsfig}.
Searches for $\nu_\mu \leftrightarrow \nu_e$ transitions with values of $L/E$ corresponding
to the atmospheric-neutrino scale are sensitive to the third mixing angle $\theta_{13}$. 
To date these searches have not observed this transition, and hence we have only an 
upper limit on $\theta_{13}$, which comes predominantly from the CHOOZ reactor experiment\cite{Apollonio:1999ae}, and is shown in the right panel of Fig.~\ref{parsfig}.

The mixing angles tell us about the flavor content of the neutrino mass eigenstates. 
Our knowledge of the $\Delta m^2_{ij}$ and the flavor content of the mass eigenstates 
is summarized in Fig.~\ref{hierfig}. Note that there are two possible patterns of 
neutrino mass. 
This is because the neutrino oscillation experiments to date have been sensitive 
to the magnitude of $\Delta m^2_{32}$, but not its sign. The neutrino spectrum shown 
on the left in Fig.~\ref{hierfig} is called the {\it Normal Mass Hierarchy} and 
corresponds to $\Delta m^2_{32} > 0$.  The neutrino spectrum shown 
on the right is called the {\it Inverted Mass Hierarchy} and corresponds to 
$\Delta m^2_{32} < 0$. The reason we don't know the sign of $\Delta m^2_{32}$, and 
hence the neutrino mass hierarchy, is that neutrino oscillations in vacuum depend 
only on the magnitude of $\Delta m^2_{32}$. However, in matter the effective parameters describing 
neutrino transitions involving electron-type neutrinos are modified in a way that is 
sensitive to the sign of $\Delta m^2_{32}$. An experiment with a sufficiently long 
baseline in matter and an appropriate $L/E$ can therefore determine the neutrino 
mass hierarchy. 

Finally, it should be noted that there is a possible complication to the simple 
three-flavor neutrino oscillation picture. 
The LSND~\cite{Aguilar:2001ty} experiment  has reported evidence for muon 
anti-neutrino to electron anti-neutrino transitions for values of $L/E$ which 
are two orders of magnitude smaller than the corresponding values observed for 
atmospheric neutrinos. The associated transition probability is very small, 
of the order of 0.3\%.
If this result is confirmed by the MiniBooNE~\cite{Church:1997ry} 
experiment, it will require a third characteristic $L/E$ range for neutrino 
flavor transitions. Since each $L/E$ range implies a different mass-splitting 
between the participating neutrino mass eigenstates, confirmation of the LSND 
result would require more than three mass eigenstates. This would be an exciting 
and radical development. Independent of whether the LSND result is confirmed 
or not, it is important that the future global neutrino oscillation program is 
able to make further tests of the three-flavor oscillation framework.

In summary, to complete our knowledge of the neutrino mixing matrix and the pattern 
of neutrino masses we must measure $\theta_{13}$ and $\delta$, determine the sign of 
$\Delta m^2_{32}$, and test the three-flavor mixing framework. The primary, initial 
goal for a Fermilab Proton Driver will be to make these measurements.

\subsection{The Importance of the Unanswered Questions}

Non-zero neutrino masses require physics beyond the SM.
The determination of neutrino masses and mixing will discriminate between various neutrino models and yield clues that will help determine whether physics beyond the SM is described by a GUT or some other theoretical framework.

The basic neutrino questions that we would like to address with a Fermilab Proton Driver are:
\begin{description}
\item{\bf What is the order of magnitude of $\theta_{13}$? }
\item{\bf Is the mass hierarchy normal or inverted? }
\item{\bf Is there CP violation in the neutrino sector and what is the value of $\delta$?}
\end{description}
These questions are discussed in the following sections.

\begin{table}[t]
\centering
\begin{tabular}{lcr}
\hline
Model(s) & Refs.\ & $\sin^2 2\theta_{13}$ \\
\hline
\hline
Minimal SO(10) & \cite{Goh:2003hf} & 0.13 \\
Orbifold SO(10) & \cite{Asaka:2003iy} & 0.04 \\
SO(10) + Flavor symmetry
 & \cite{Babu:1998wi} & $1.2 \cdot 10^{-6}$ \\
 & \cite{Albright:2001uh} & $7.8 \cdot 10^{-4}$ \\
 & \cite{Blazek:1999hz,Ross:2002fb,Raby:2003ay} &
   0.01 .. 0.04 \\
 & \cite{Kitano:2000xk,Maekawa:2001uk,Chen:2002pa} &
   0.09 .. 0.18 \\
SO(10) + Texture
 & \cite{Bando:2003ei} & $4 \cdot 10^{-4}$ .. 0.01 \\
 & \cite{Buchmuller:2001dc} & 0.04 \\
$\mathrm{SU}(2)_\mathrm{L} \times \mathrm{SU}(2)_\mathrm{R} \times \mathrm{SU}(4)_c$ &
 \cite{Frampton:2004vw} & 0.09 \\
\hline
Flavor symmetries
 & \cite{Grimus:2001ex,Grimus:2003kq,Grimus:2004rj} & 0 \\
 & \cite{Chen:2004rr,Aizawa:2004qf,Mohapatra:2004mf} &
   $\lesssim 0.004$ \\
 & \cite{Antusch:2004xd,Antusch:2004re,Rodejohann:2004qh} & $10^{-4}$ .. 0.02 \\
 & \cite{Babu:2002dz,Ohlsson:2002rb,King:2003rf,Shafi:2004jy,Mohapatra:2004mf} & 0.04 .. 0.15 \\
\hline
Textures
 & \cite{Bando:2003wb} & $4 \cdot 10^{-4}$ .. 0.01 \\
 & \cite{Honda:2003pg,Lebed:2003sj,Ibarra:2003xp,Harrison:2004he} & 0.03 .. 0.15 \\
\hline
$3 \times 2$ see-saw & \cite{Frampton:2002qc} & 0.04 \\
 & \cite{Mei:2003gn} (n.h.) & 0.02 \\
 & \hphantom{\cite{Mei:2003gn}}(i.h.) &
 $> 1.6 \cdot 10^{-4}$ \\
\hline
Anarchy & \cite{deGouvea:2003xe} & $>0.04$ \\
\hline
Renormalization group enhancement & \cite{Mohapatra:2003tw} & 0.03 .. 0.04 \\
\hline
M-Theory model & \cite{Arnowitt:2003kc} & $10^{-4}$ \\
\hline
\end{tabular}
\caption{
 Selection of predictions for $\sin^2 2\theta_{13}$.
 The numbers should be considered as order of magnitude statements. The
 abbreviations ``n.h.'' and ``i.h.'' refer to the normal and inverted hierarchies, respectively.
}
\label{tab:PredictionsT13}
\end{table}

\subsubsection{The importance of $\theta_{13}$}

Neutrino oscillation experiments have shown that two of the neutrino 
mixing angles ($\theta_{23}$ and $\theta_{12}$) are large. This was a 
surprise since the corresponding mixing angles in the quark mixing matrix 
are small. We have only an upper limit on the third neutrino mixing 
angle $\theta_{13}$. From this limit we know that $\theta_{13}$ is much 
smaller than $\theta_{23}$ or $\theta_{12}$. However we have no good 
reason to expect it to be very small. Predictions from recent models 
are listed in Table~\ref{tab:PredictionsT13}. Most of the presently 
viable neutrino mass models predict that $\theta_{13}$ is close to the 
present bound. A value of $\theta_{13}$ much smaller than the bound 
would suggest a new flavor symmetry that suppresses this mixing. 
However, even if $\theta_{13}$ is 
exactly zero at the GUT scale, radiative corrections would be expected 
to drive its value away from zero at laboratory energies.
In any case 
{\it determining the order of magnitude of $\theta_{13}$ will discriminate between theoretical models (Table~\ref{tab:PredictionsT13}) and provide 
crucial guidance toward an understanding of the physics of neutrino masses}.

In addition to its theoretical importance, the size of 
$\theta_{13}$ has important experimental consequences. If $\theta_{13}$ 
is within an order of magnitude of its present bound we will probably know its value 
before the ``Proton Driver Era''. A Fermilab Proton Driver would then enable the mass 
hierarchy to be determined and a search for CP violation to be made. If 
$\theta_{13}$ is small ($\sin^22\theta_{13}<0.01$) 
we will not know its value before the Proton Driver Era. 
The initial Fermilab Proton Driver program would then improve our knowledge of $\theta_{13}$ and prepare the way for a second generation program. If $\sin^22\theta_{13} \gtrsim 0.005$ the initial Fermilab Proton Driver experiment would establish its value and might also determine the mass hierarchy, but would not be sufficiently sensitive to search for CP violation. A second generation program will be required. The options for this second generation include an upgraded detector with or without a new beamline, and a neutrino factory driven by the Proton Driver.
 Note that {\it the value of $\theta_{13}$ will determine which 
facilities and experiments will be needed beyond the initial Fermilab Proton Driver 
experiments to complete the neutrino oscillation program}.

\subsubsection{The importance of the Mass Hierarchy}

Specific neutrino mass models are usually only compatible with one of the 
two possible neutrino mass hierarchies (normal or  inverted). A measurement 
of the sign of $\Delta m^2_{31}$ would therefore discriminate between models. 
For example, GUT models with a standard type~I see-saw mechanism 
tend to predict a normal hierarchy (see, for example, the reviews 
Refs~\cite{Chen:2003zv,Altarelli:2004za}),
while an inverted hierarchy is often 
obtained in models that employ flavor symmetries such as 
$L_e-L_\mu-L_\tau$~\cite{Petcov:1982ya,Leung:1983ti}.

A determination of the sign of $\Delta m^2_{31}$ would also have some 
consequences for neutrinoless double beta decay experiments. A negative 
$\Delta m^2_{31}$ would imply a lower limit on the effective mass for 
neutrinoless double beta decay (in the case of Majorana neutrinos) 
which would be expected to be within reach of the next generation of 
experiments.

\subsubsection{The importance of CP Violation and $\delta$}

Leptogenesis~\cite{Fukugita:1986hr}, in which CP violation in the leptonic 
Yukawa couplings ultimately results in a baryon asymmetry in the early 
Universe, provides an attractive possible explanation for the observed 
baryon asymmetry. In the most general case, the CP phases involved in 
leptogenesis are not related to the low-energy CP phases that appear in the 
effective neutrino mass matrix~\cite{Branco:2001pq,Pascoli:2003uh}. However, 
specific neutrino mass models can yield relationships between the CP violation 
relevant to leptogenesis 
and the low-energy CP phases. Indeed, to successfully obtain leptogenesis some 
models require a non-zero phase $\delta$ (see, for example, 
Refs~\cite{Frampton:2002qc,King:2002qh}). 
Hence, although a measurement of CP violation in the neutrino sector would 
not establish leptogenesis as the right explanation for the observed baryon 
asymmetry, it would be suggestive and a measurement of $\delta$ would discriminate between explicit neutrino mass models.

\subsubsection{Other Oscillation Physics}

With a Proton Driver the Fermilab neutrino program would provide a path to 
the ultimate sensitivity for measurements of $\theta_{13}$, the mass hierarchy, 
and CP violation. In addition to these crucial measurements, to discriminate 
between different theoretical models, it will also be important to improve 
the precision of the other oscillation parameters ($\theta_{12}, \theta_{23}, 
\Delta m^2_{21}, \Delta m^2_{32}$). Note that $\theta_{23}$ is of particular 
interest as its current, poorly determined, value is consistent with maximal 
mixing in the (2,3) sector. Is this mixing really maximal? Furthermore the 
level of consistency between the precisely measured values of the parameters 
in the various appearance and disappearance modes will test the 3 flavor 
mixing framework, possibly leading to further exciting discoveries. The 
Fermilab Proton Driver would not only address the value of $\theta_{13}$, the 
mass hierarchy, and CP violation, but would also provide a more comprehensive 
set of measurements that could lead to further unexpected surprises.

\subsection{Evolution of the Sensitivity to $\theta_{13}$}

In the coming years we can expect improvements in our knowledge of the 
oscillation parameters from the present generation of running experiments 
(MiniBooNE, KamLAND, K2K, MINOS, SNO, SuperK) and experiments under construction (T2K). 
Beyond this a new generation of reactor and accelerator based experiments 
are being proposed (for example the Double-CHOOZ reactor experiment and the 
NO$\nu$A experiment using the Fermilab NuMI beam). 
In the coming decade the search for a non-zero 
$\theta_{13}$ is of particular importance since, not only is $\theta_{13}$ the only unmeasured mixing angle, but its value will determine the prospects for determining the mass hierarchy and making a sensitive search for CP violation.

\begin{figure}[t!]
\centering \includegraphics[width=10cm]{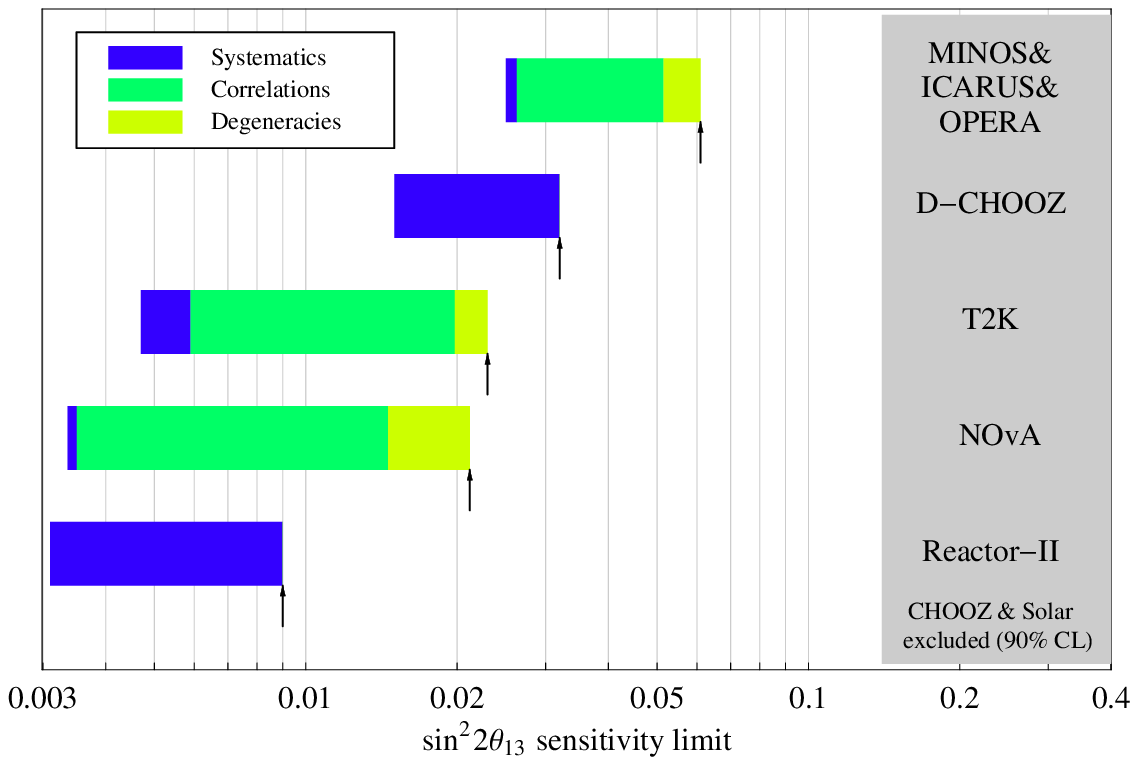}
   \caption{\label{fig:externcomp} The $\stheta$ sensitivity limit
   at the 90\% CL for \minos, \icarus, and \opera\ combined, \DChooz,
   \JHFSK, 2$^{\rm nd}$ generation reactor experiment (\ReactorII\ ),
   and NO$\nu$A. The left edges of the bars are obtained for the
   statistics limits only, whereas the right edges are obtained after
   successively switching on systematics, correlations, and
   degeneracies, \ie, they correspond to the final $\stheta$ sensitivity
   limits. The gray-shaded region corresponds to the current
   $\sin^22\theta_{13}$ bound at 90\% CL. For the true values of the
   oscillation parameters, we use $|\Delta m_{31}^2| = 2.0 \cdot 10^{-3}\,\mathrm{eV}^2$,
   $\sin^22\theta_{23} = 1$, $\Delta m_{21}^2 = 7.0 \cdot 10^{-5}\,\mathrm{eV}^2$, $\sin^22\theta_{12} = 0.8$~\cite{Fogli:2003th,Maltoni:2003da,Ahmed:2003kj,SKupdate}, and a normal mass hierarchy. Figure extended version from \Ref~\cite{Huber:2004ug}.}
\end{figure}

In this section we describe the expected evolution of our sensitivity to 
$\theta_{13}$ over the next ten to fifteen years, a time period that includes a first generation of Fermilab Proton Driver experiments. A list of relevant experiments and 
their characteristics is given in Table~\ref{tab:reps}. The $\sin^2 2\theta_{13}$ 
sensitivities for these experiments are summarized in Fig.~\ref{fig:externcomp}, 
where the other oscillation parameters have been chosen to correspond to the 
present central values. Note that in general the sensitivities of the accelerator based experiments are dominated by statistical uncertainties. To make further progress will require larger detectors and higher intensity beams. Reactor-based experiments, by contrast, are dominated by systematic uncertainties.

In interpreting Fig.~\ref{fig:externcomp} it is important to note that it shows the {\em 90\% sensitivity} in contrast to the {\em $3\sigma$ discovery reach} shown in Figs.~\ref{fig:timescale_small} and \ref{fig:timescale}. The 90\% sensitivity is calculated by setting the ``true'' value of $\sin^2 2\theta_{13}$ equal to zero in the calculation and then finding the limit on $\sin^2 2\theta_{13}$ that can be set at 90\% confidence given all possible values of $\delta$ and the mass hierarchy. It is the appropriate quantity to calculate if one wants to understand how low an experiment can set a limit on the value of $\sin^2 2\theta_{13}$. The $3\sigma$ discovery reach is calculated by finding the value of $\sin^2 2\theta_{13}$ that can be distinguished from zero at $3\sigma$ confidence. It is the appropriate quantity if one is interested in knowing when one can determine that $\sin^2 2\theta_{13}$ is non-zero. Because of the correlated effects of $\sin^2 2\theta_{13}$, $\delta$ and the hierarchy (among others) we can determine that $\sin^2 2\theta_{13}$ is non-zero well before measuring its actual value. From the perspective of a Fermilab Proton Driver program the $3\sigma$ discovery reach is the more relevant quantity as the interest is in knowing that $\sin^2 2\theta_{13}$ is non-zero. Then one can set about designing an optimised program to go after CP violation and the mass hierarchy.

\begin{table}[t]
{\small
\begin{center}
\begin{tabular}{lrrrlrr}
\hline
Label & $L$ & $\langle E_\nu \rangle$ & $P_{\mathrm{Source}}$ &
Detector technology & $m_{\mathrm{Det}}$ & $t_{\mathrm{run}}$ \\
\hline
\multicolumn{7}{l}{\bf{Conventional beam experiments:}} \\
\minos\ & $735 \, \mathrm{km}$ & $3 \,\mathrm{GeV}$ &
$3.7 \cdot 10^{20} \,\mathrm{pot/y}$ &
Magn. iron calorim. &  $5.4\,\mathrm{kt}$ & $5 \, \mathrm{yr}$ \\
\icarus\ & $732\,\mathrm{km}$ &  $17\,\mathrm{GeV}$  &
$4.5 \cdot 10^{19}\,\mathrm{pot/y}$ &
Liquid Argon TPC & $2.35\,\mathrm{kt}$ & $5 \, \mathrm{yr}$\\
\opera\ & $732\,\mathrm{km}$ &  $17\,\mathrm{GeV}$ &
$4.5 \cdot 10^{19}\,\mathrm{pot/y}$ &
Emul. cloud chamb. &  $1.65\,\mathrm{kt}$ & $5 \, \mathrm{yr}$\\[0.1cm]
\multicolumn{7}{l}{\bf{Off Axis experiments:}} \\
\JHFSK\ & $295  \, \mathrm{km}$ & $0.76 \, \mathrm{GeV}$ &
$1.0 \cdot 10^{21} \, \mathrm{pot/y}$  &
Water Cherenkov & $22.5 \, \mathrm{kt}$ & $5 \, \mathrm{yr}$ \\
NO$\nu$A$^\dag$ & $810 \, \mathrm{km}$ & $2.22 \, \mathrm{GeV}$ &
$6.5 \cdot 10^{20} \,\mathrm{pot/y}$ &
TASD & $30 \, \mathrm{kt}$ & $5 \, \mathrm{yr}$ \\[0.1cm]
\multicolumn{7}{l}{\bf{Reactor experiments:}} \\
\CHOOZII$^\dag$\ & $1.05 \, \mathrm{km}$ & $\sim 4 \, \mathrm{MeV}$ & $2
\times 4.25 \, \mathrm{GW}$ &
Liquid Scintillator & $11.3 \, \mathrm{t}$ & $3 \, \mathrm{yr}$ \\
\ReactorII$^\dag$\ & $1.70 \, \mathrm{km}$ & $\sim 4 \, \mathrm{MeV}$ &
$ 8\,\mathrm{GW}$ &
Liquid Scintillator & $200\,\mathrm{t}$ & $5\,\mathrm{yr}$ \\
\hline
$\dag$ proposed \\
\end{tabular}
\end{center}
}
\caption{\label{tab:reps} The different experiments discussed in the text.
   The table shows, for each experiment,
   the baseline $L$, the mean neutrino energy $\langle E_\nu \rangle$,
   the source power $P_{\mathrm{Source}}$ (for beams: in protons on
   target per year, for reactors: in gigawatts of thermal reactor
   power), the detector technology, the fiducial detector mass
   $m_{\mathrm{Det}}$, and the running time $t_{\mathrm{run}}$. Note
   that most results are, to a first approximation, a function of the
   product of running time, detector mass, and source power.
   Table modified from Ref~\cite{Huber:2004ug}.}
\end{table}

\subsubsection{Conventional Beam Experiments}

MINOS is a muon-neutrino disappearance experiment that is expected to confirm 
the oscillation interpretation of the atmospheric neutrino data and to 
better determine the associated $|\Delta m^2|$. MINOS will also have 
some capability to detect electron-neutrino appearance, and hence has 
some sensitivity to $\theta_{13}$. However, this sensitivity is limited. 
MINOS is just beginning to take data. In the coming 5 years we 
expect MINOS to determine $\theta_{13}$ if it is very close to the present 
bound. MINOS is also expected to reduce the uncertainty on 
$|\Delta m^2_{31}|$ to about $\pm 10\%$.

In Europe, the CNGS program consists of two experiments, ICARUS and OPERA, 
designed to study tau-neutrino appearance with an L/E corresponding to 
the atmospheric neutrino oscillation scale. They will also have
sensitivity to electron-neutrino appearance, and hence to $\theta_{13}$. 
The CNGS experiments are expected to begin running in a few years. 
After 5 years 
of data taking, the combined sensitivity of ICARUS, OPERA, and MINOS will 
enable $\sin^2 2\theta_{13}$ to be determined if it exceeds $\sim 0.06$.

\subsubsection{Off-Axis Experiments}

Looking further into the future, significant progress could be made with 
a new long-baseline experiment that exploits the NuMI beamline, together 
with a Proton Driver, and a detector that is optimized 
to detect $\nu_e$ appearance. Although no NuMI upgraded experiment has yet 
received final approval, we might 
imagine that NO$\nu$A, or an equivalent experiment, is approved, constructed, 
and becomes operational in 5-10 years from now. After 5 years of data 
taking we would expect NO$\nu$A to determine $\sin^2 2\theta_{13}$ if it 
exceeds $\sim 0.02$. 

In Japan the JPARC beamline for T2K, a high-statistics, off-axis, second generation version 
of K2K, has been approved, and is expected to be completed in 2009. With 5 
years of data taking T2K is expected to determine $\sin^2 2\theta_{13}$ if 
it exceeds $\sim 0.02$. The combined NO$\nu$A and T2K sensitivity would be in 
the range 0.01-0.02.

\subsubsection{Reactor Experiments}

A new generation of reactor experiments are being proposed with detectors 
and baselines chosen to be sensitive to $\theta_{13}$. Although choices 
still have to be made to determine which of these experiments are supported, 
it seems reasonable to assume that one or two reactor experiments will be 
executed in the coming decade. The Double-CHOOZ experiment appears to be 
furthest along in the approval process. This experiment is expected to 
determine $\sin^2 2\theta_{13}$ if it exceeds $\sim 0.03$. Beyond this, 
a more ambitious reactor experiment, referred to in Table \ref{tab:reps} as Reactor II,
 might be expected to reach a 
sensitivity $\sin^2 2\theta_{13}$ approaching 0.01 at 90\% confidence (the curves in Figs.~\ref{fig:timescale_small} and \ref{fig:timescale} are for 3$\sigma$ discovery).
This sensitivity is limited by systematic uncertainties but, if achieved, will be slightly better, for some values of $\delta$, than the corresponding sensitivity expected for T2K or NO$\nu$A, but worse for other values of $\delta$.

\subsubsection{The Evolution}

\begin{figure}[t!]
   \centering \includegraphics[width=\textwidth]{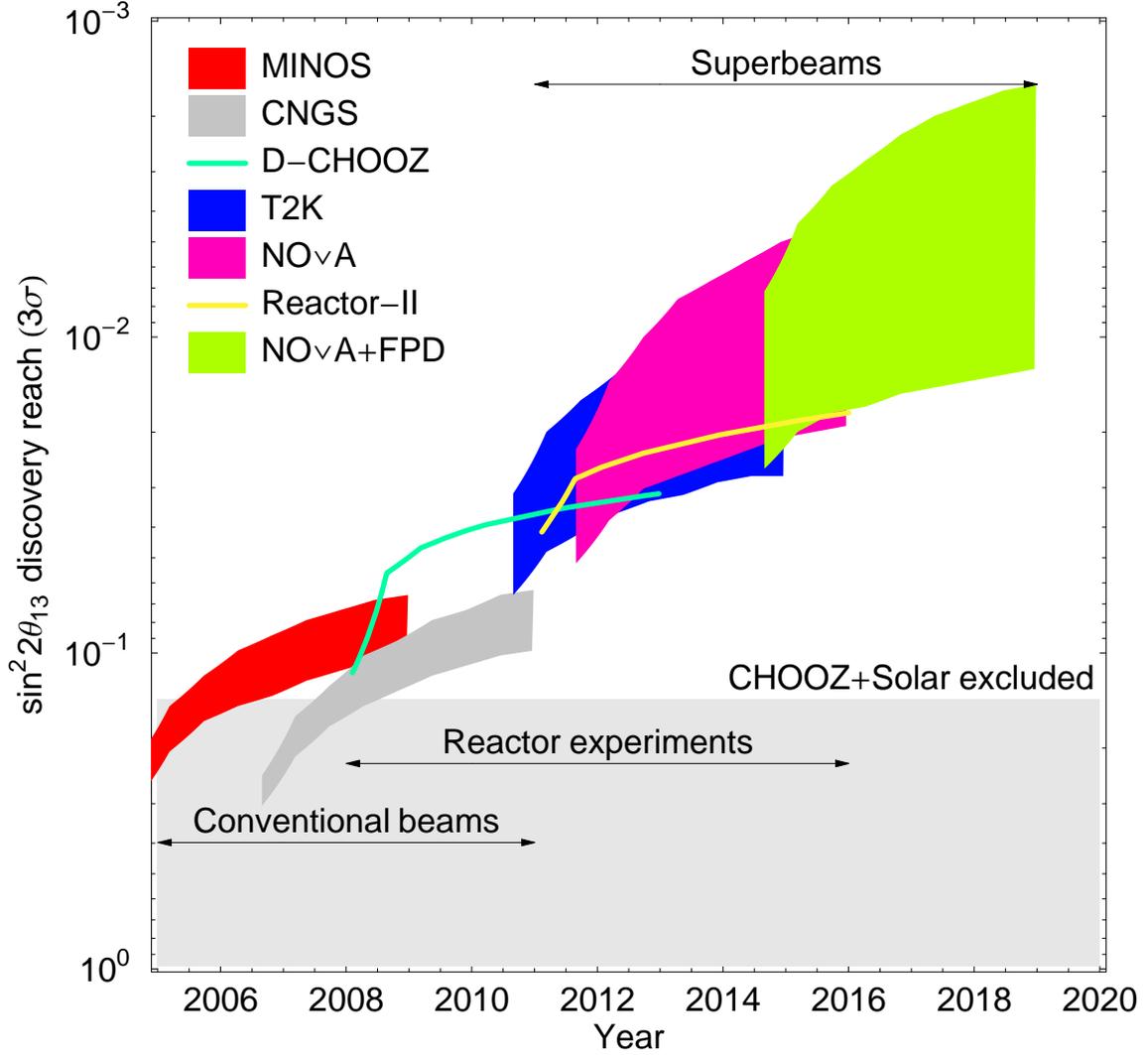}
   \vspace*{0.1cm}
   \caption{ \label{fig:timescale_small} Anticipated evolution of the $\stheta$ discovery reach for the global neutrino program. The curves show the 3$\sigma$ sensitivities for each experiment to observe a non-zero value of $\sin^2 2\theta_{13}$. The bands reflect the dependence of the sensitivity on the CP violating phase $\delta$. The calculations are based on the experiment simulations in \Refs~\cite{Huber:2002mx,Huber:2004ug} and include statistical and systematic uncertainties and parameter correlations. They assume a normal hierarchy and $\ldm = 2.5 \cdot 10^{-3} \, \mathrm{eV}^2$, $\sin^2 2 \theta_{23}=1$, $\sdm = 8.2 \cdot 10^{-5} \, \mathrm{eV}^2$, $\sin^2 2 \theta_{12} = 0.83$. All experiments are operated with neutrino running only and the full detector mass is assumed to be available right from the beginning. The starting times of the experiments have been chosen as close as possible to those stated in the respective LOIs. \ReactorII\ and FPD refer, respectively, to a 2$^{\rm nd}$ generation reactor experiment and to the Fermilab Proton Driver.
   }
\end{figure}

\begin{figure}[t!]
   \centering \includegraphics[width=\textwidth]{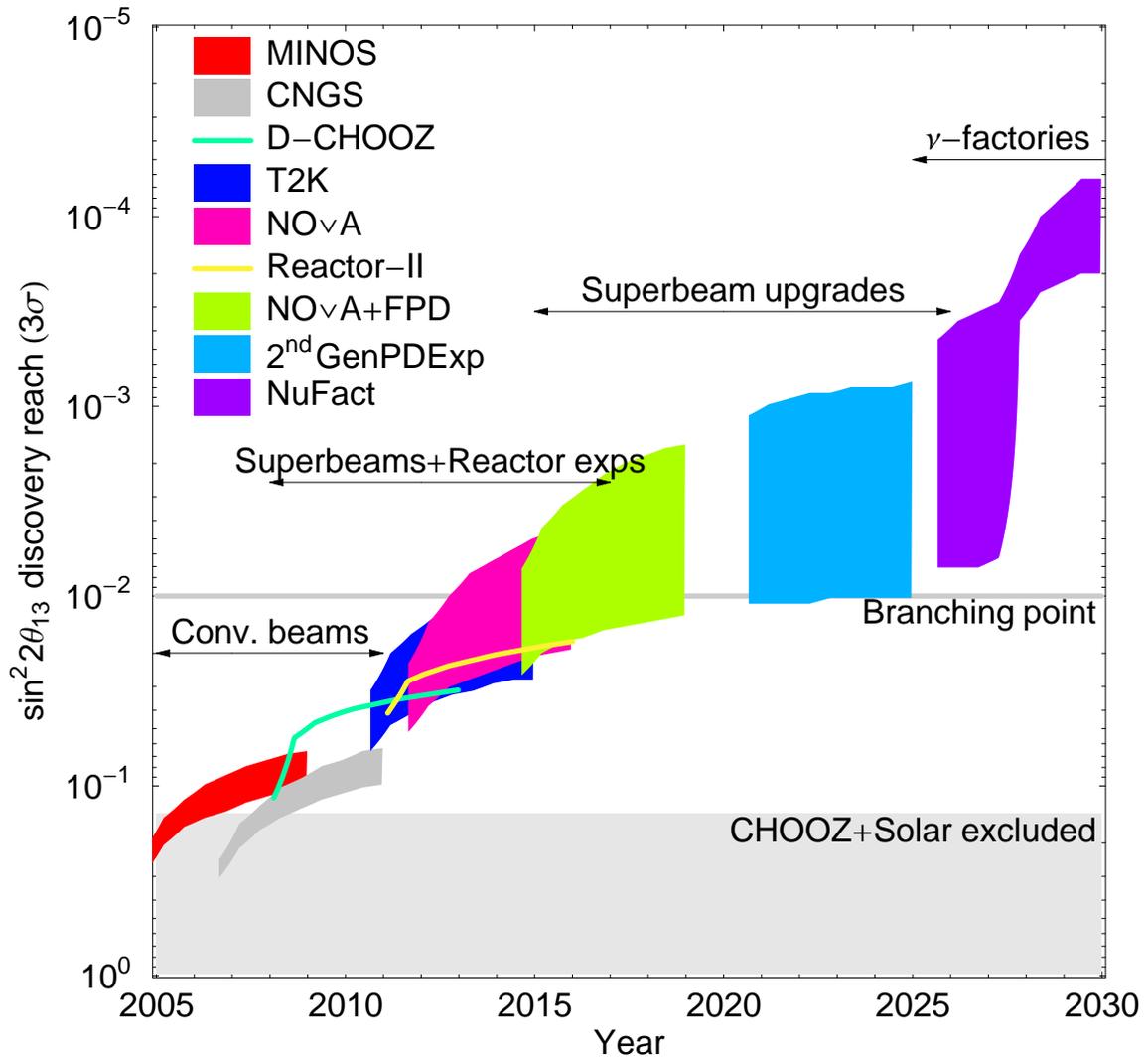}
   \vspace*{0.1cm}
   \caption{\label{fig:timescale} Anticipated evolution of the $\sin^2 2\theta_{13}$ discovery reach for the global neutrino program. See caption for Fig.~\ref{fig:timescale_small} for details. The ``branching point'' refers to the decision point for the experimental program (rather than the Fermilab Proton Driver), \ie, between an upgraded beam and/or detector and a neutrino factory  program. The upgrade 2ndGenPDExp (Second Generation Proton Driver Experiment) is assumed to start ten years after \JHFSK\ starts and the curve uses numbers from the T2HK proposal. The neutrino factory is assumed to start about ten years after the branching point and to switch polarity after 2.5 years. 
   }
\end{figure}

The anticipated evolution of the $\sin^2 2\theta_{13}$ discovery reach of the 
global neutrino oscillation program is illustrated in Fig.~\ref{fig:timescale_small}. 
The sensitivity is expected to improve by about an order of magnitude over the 
next decade. This progress is expected to be accomplished in several steps, 
each yielding a factor of a few increased sensitivity. 
During this first decade the Fermilab program will have contributed to the 
improving global sensitivity with MINOS, followed by NO$\nu$A. MINOS is the on-ramp 
for the US long-baseline neutrino oscillation program. NO$\nu$A would be the next step. 
Note that we assume that NO$\nu$A starts taking data with the existing beamline before 
the Proton Driver era. The Proton Driver would take NO$\nu$A into the fast lane of the 
global program. Also note that the accelerator based and reactor based experiments are complementary. In particular, the reactor experiments make disappearance measurements, limited by systematic uncertainties. The NO$\nu$A experiment is an appearance experiment, limited by statistical uncertainties, and probes regions of parameter space beyond the reach of the proposed reactor experiments.

\clearpage

\clearpage

\subsection{Fermilab Proton Driver Oscillation Physics Program}

Independent of the value of $\theta_{13}$ the initial Fermilab Proton 
Driver long-baseline neutrino experiment (NO$\nu$A+FPD) is expected to make an 
important contribution to the global oscillation program. 
If $\theta_{13}$ is very small NO$\nu$A+FPD would be expected to provide the 
most stringent limit on this important parameter, and prepare the 
way for a neutrino factory. If $\theta_{13}$ is sufficiently large, 
NO$\nu$A+FPD would be expected to measure its value, perhaps determine the mass 
hierarchy, and prepare the way for a sensitive search for CP violation.  
The evolution of the Fermilab Proton Driver physics program beyond the 
initial experiments will depend not only 
on $\theta_{13}$, but also on what other neutrino experiments will be 
built elsewhere in the world. In considering the long-term evolution 
of the Fermilab Proton Driver program we must take into account the 
uncertainty on the magnitude of $\theta_{13}$ and consider how the 
global program might evolve.

We begin by considering the evolution of the program if 
$\sin^2 2\theta_{13} < 0.01$. In this case we will know that ultimately we 
will need a neutrino factory to complete all of the important 
oscillation measurements. The initial Fermilab Proton Driver 
experiment would be a search experiment that would improve the 
limit on, or establish the value of, $\theta_{13}$.  
Fig.~\ref{fig:timescale} shows a longer-term version of the $\sin^2 2\theta_{13}$ discovery reach 
versus time plots shown in Fig.~\ref{fig:timescale_small}. The initial Fermilab Proton Driver experiment would begin to explore 
the region below $\sin^2 2\theta_{13} \sim 0.01$ and could be upgraded 
to further improve the sensitivity by a factor of a few. The neutrino 
factory would ultimately provide a two orders of magnitude improvement 
in sensitivity.

We have no reason to expect a very small value for $\theta_{13}$. 
Hence, as the global program achieves increasing sensitivity to 
$\theta_{13}$, at any time a finite value might be established, 
and the focus of the program will change from searching for evidence 
for $\nu_\mu \leftrightarrow \nu_e$ transitions to measuring $\theta_{13}$, 
determining the mass hierarchy, and searching for CP violation. 
To explore how in this case the Fermilab Proton Driver would contribute 
to the global program we consider four cases:
\begin{description}
\item[{\bf Case 1:}] No Fermilab Proton Driver and no upgrade to the T2K beam.
\item[{\bf Case 2:}] An upgrade to the T2K beam, but no Fermilab Proton Driver.
\item[{\bf Case 3:}] A Fermilab Proton Driver, but no upgrade to the T2K beam.
\item[{\bf Case 4:}] Both a Fermilab Proton Driver and an upgraded T2K beam.
\end{description}

%
\begin{figure}[!ht]
\includegraphics[width=\textwidth]{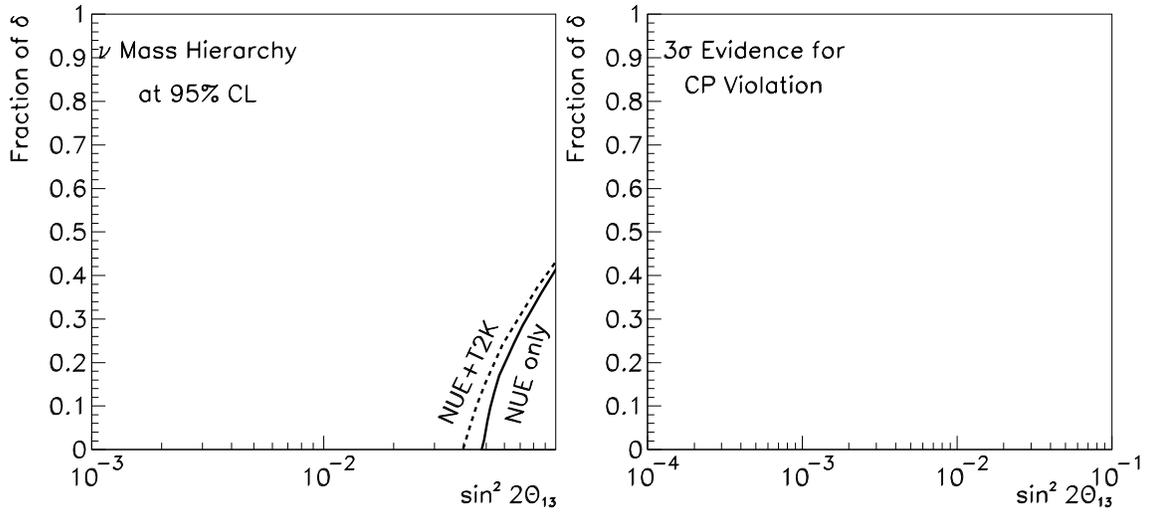}
\caption{{\bf Case 1: No Fermilab Proton Driver and no upgrade to the T2K beam.} Regions of parameter space where the mass hierarchy (left) and CP 
violation (right) can be observed at 95\% CL and at 3$\sigma$, 
respectively.  Note that CP violation would not be visible at all and T2K alone is not sensitive to the hierarchy. NO$\nu$A, T2K, etc are defined in Table~\ref{tab:summary1}.\label{fig:reach_no_pd}
\vspace*{2.9cm}}
\end{figure}

\begin{figure}[!hb]
\includegraphics[width=\textwidth]{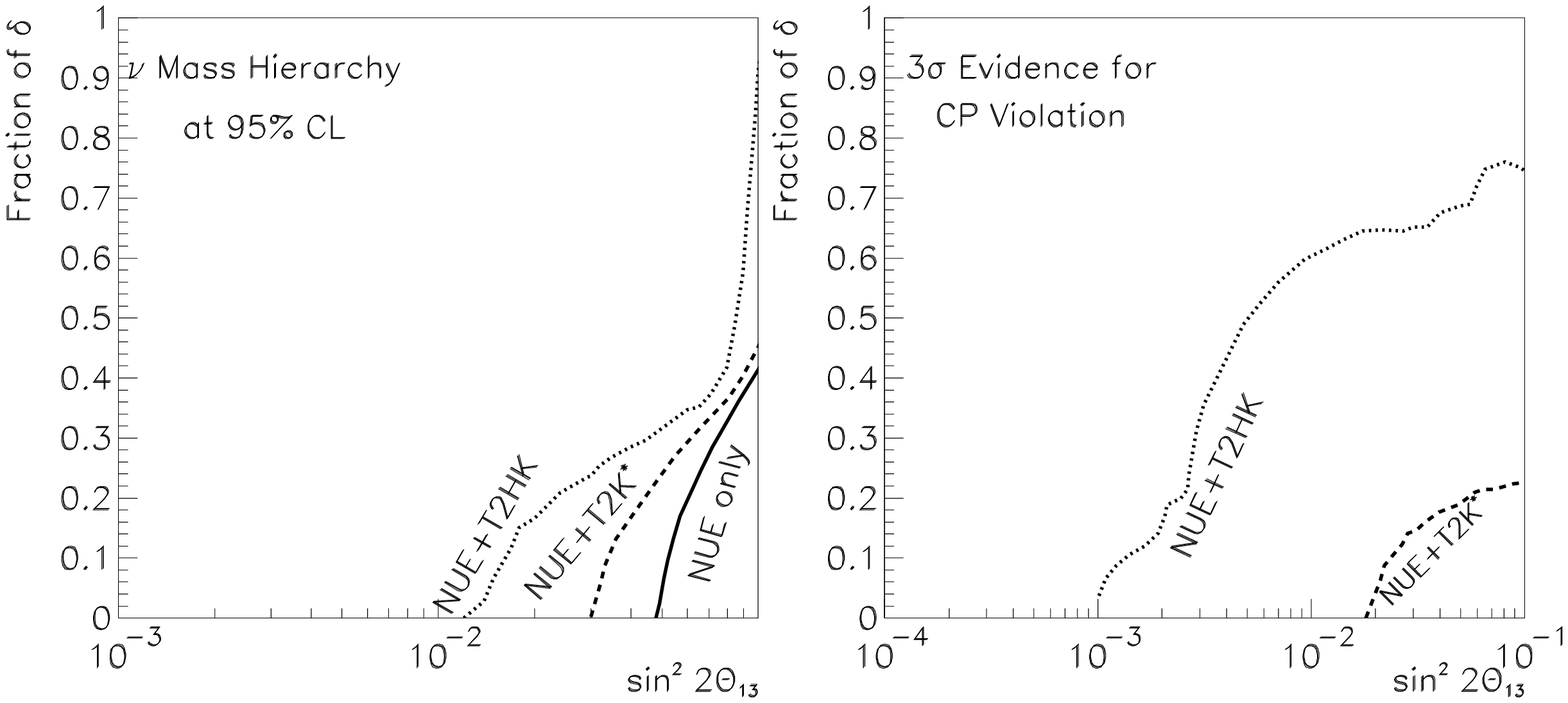}
\caption{{\bf Case 2: An upgrade to the T2K beam, but no Fermilab Proton Driver.} Regions of parameter space where the mass hierarchy (left) and CP 
violation (right) can be observed  at 95\% CL and at 3$\sigma$, 
respectively. NO$\nu$A, T2K, etc are defined in Table~\ref{tab:summary1}.\label{fig:reach_japan_pd}}
\end{figure}

\begin{figure}[!ht]
\includegraphics[width=\textwidth]{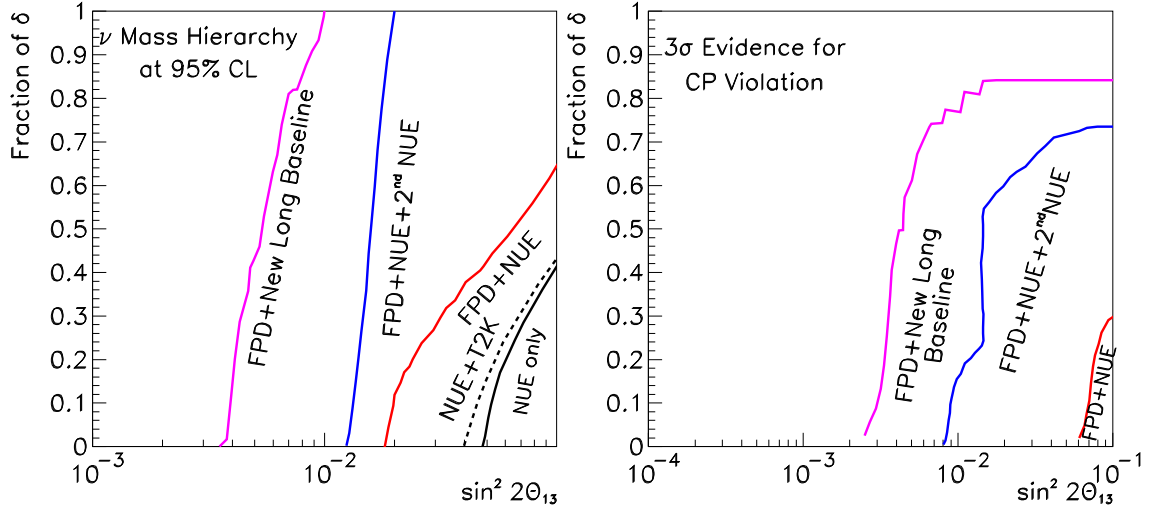}
\caption{{\bf Case 3: A Fermilab Proton Driver, but no upgrade to the T2K beam.} Regions of parameter space where the mass hierarchy (left) and CP 
violation (right) can be observed  at 95\% CL and at 3$\sigma$, 
respectively. NO$\nu$A, T2K, etc are defined in Table~\ref{tab:summary1}. The addition of T2K data would not significantly change the position of these curves.\label{fig:reach_fnal_pd}
\vspace*{2.8cm}}
\end{figure}

\begin{figure}[!hb]
\includegraphics[width=\textwidth]{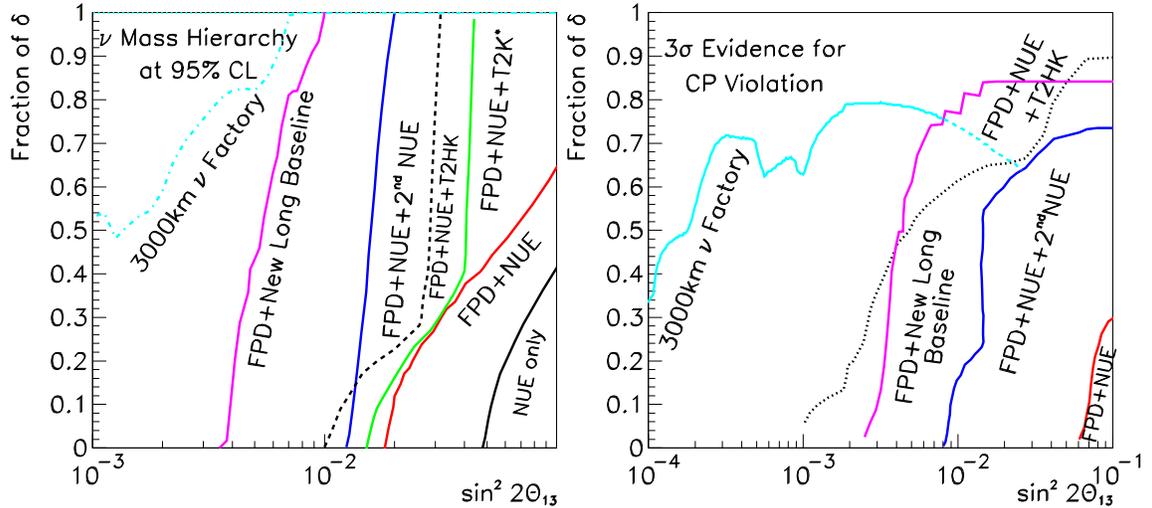}
\caption{{\bf Case 4: Both a Fermilab Proton Driver and an upgraded T2K beam.} Regions of parameter space where the mass hierarchy (left) and CP 
violation (right) can be observed at 95\% CL and at 3$\sigma$, 
respectively. NO$\nu$A, T2K, etc are defined in Table~\ref{tab:summary1}.\label{fig:reach_all_pd}}
\end{figure}

\begin{table}[ht]
\begin{center}
\begin{tabular}{|l|c|c|c|} \hline
     & Detector Mass & Proton Power & Running Time \\ 
Name &   (kton)      &   (MW)       &  (years)      \\ \hline 
NO$\nu$A & 30 & 0.65 & 3$\nu$ + 3$\bar\nu$ \\ \hline 
FPD+NO$\nu$A & 30 & 2.0 & 3$\nu$ + 3$\bar\nu$ \\ \hline 
FPD+NO$\nu$A+$2^{nd}$ NO$\nu$A & 30(+30)  & 2 &  6 (3)$\nu$ + 6 (3) $\bar\nu$ \\ \hline
FPD+New Long Baseline &  125 or 500 & 2 + 2 &  5$\nu$ + 5$\bar\nu$ \\ \hline
T2K   &   50 &  0.77 &  3$\nu$ + 3$\bar\nu$ \\ \hline
T2K$^*$ &  50 & 4 &   3$\nu$ + 3$\bar\nu$ \\ \hline
T2HK   &  500 & 4  &  3$\nu$ + 3$\bar\nu$ \\ \hline
3000km $\nu$ Factory & 50 & 4 & 3$\nu$ + 3$\bar\nu$ \\ \hline
\end{tabular} 
\caption{Summary of the various experiments which are discussed in the text and Figs.~\ref{fig:timescale_small} and \ref{fig:timescale}.}
\label{tab:summary1} 
\end{center}
\end{table}

The prospects for determining the neutrino mass hierarchy and discovering CP violation 
depend upon the values of $\theta_{13}$ and $\delta$.
Figures~\ref{fig:reach_no_pd},\ref{fig:reach_japan_pd}, \ref{fig:reach_fnal_pd} and 
\ref{fig:reach_all_pd} show as a function of 
$\sin^2 2\theta_{13}$, for various combinations of experiments,  
the fraction of all possible values of $\delta$ for which the 
mass hierarchy can be determined (left panels) and CP violation can be 
discovered (right panels). 

The first case, where there is no Fermilab Proton Driver and no 
upgrade to the T2K beam, is shown in Fig.~\ref{fig:reach_no_pd}.  Note that even
by combining T2K and NO$\nu$A data, one cannot arrive at a 3$\sigma$ determination
for CP violation.  If, in addition to no Fermilab Proton Driver, there is no NO$\nu$A 
then T2K alone will not be able to determine the mass hierarchy. 
NO$\nu$A would provide some sensitivity to the mass hierarchy, which would be 
somewhat improved by combining NO$\nu$A and T2K results.

The second case, where there is no Fermilab Proton Driver but the T2K beam 
is upgraded, is shown in Fig.~\ref{fig:reach_japan_pd}.  We assume that NO$\nu$A is built, 
which will provide some sensitivity to the mass hierarchy. 
In this case, there is some parameter space where CP violation can be seen
at $3\sigma$, which expands dramatically if there is a 20-fold
increase in detector mass which would happen if Hyper-Kamiokande
were to be built.  However, there would always be a significant fraction
of $\delta$ for which there would be an ambiguity due to the uncertain 
mass hierarchy, which means that a degenerate CP conserving solution may overlap the CP violating 
solution and destroy the CP violation sensitivity. Since the baseline is 295~km, substantial beam and detector upgrades to T2K make only a modest impact on the mass hierarchy sensitivity for a limited fraction of the $\delta$ parameter space. 

The third case, shown in Fig.~\ref{fig:reach_fnal_pd}, is where there
is a Fermilab Proton Driver, but no T2K upgrade  
The curves show various options:
either running with one or more detectors located at different off axis
angles from the NuMI beamline, or with a new long baseline experiment
with a new beamline. Note that the Fermilab Proton Driver yields a dramatic 
improvement in the potential to determine the mass hierarchy, which compares 
favorably with Case 2. The initial Fermilab Proton Driver experiment would 
have limited sensitivity to CP violation, but further upgrades to the beamline 
and detector would provide a significantly improved sensitivity which is again 
favorable when compared to Case 2.

The fourth case, shown in Fig.~\ref{fig:reach_all_pd}, is where there
is a both a Fermilab Proton Driver and an upgraded T2K program.  
If in the initial program the fluxes are increased, but detectors are not upgraded,
then there is some improvement in sensitivity over Case 3, particularly for the 
mass hierarchy at large $\sin^2 2\theta_{13}$.

In presenting the impact of a Fermilab Proton Driver on the global neutrino program we have featured an off-axis narrow band beam experiment, NO$\nu$A. Recently a group from Brookhaven has proposed an alternative approach which exploits an on-axis broad band beam with a long baseline (L = 2540 km corresponding to BNL to the Homestake mine) \cite{Diwan:2003bp}. To understand whether this approach could also be implemented at Fermilab calculations have been made \cite{Diwan:2004bt} for a baseline of 1290 km, corresponding to FNAL to the Homestake mine. The resulting precision in the ($\sin^2 2\theta_{13}, \delta$) plane is found to be comparable to or better than the L = 2540 km case. Whether the broadband beam concept is better or worse than the off-axis concept depends critically on the assumed background levels for the broadband experiment. A third alternative has been proposed in which a broad energy range is covered by a set of narrow band beams going to the same detector, the tighter energy spread significantly reducing backgrounds. One of these neutrino beams would be produced using the 8 GeV linac beam, and would require the highest practical primary beam power ($\sim2$ megawatts). Whichever is ultimately preferred, the Fermilab Proton Driver would be able to accommodate any of these alternatives.

In summary, although we do not know the value of $\theta_{13}$ or at what 
point in time its value will be known, we do know that the Fermilab Proton Driver 
will offer choices that will enable it to provide a critical contribution 
to the global program. In all the cases considered, without a Fermilab Proton Driver 
the sensitivity to the neutrino mass hierarchy will be very limited.

\cleardoublepage

\section{Neutrino Scattering}

While neutrino oscillation experiments probe the physics of neutrino masses 
and mixing, neutrino scattering experiments probe the interactions of 
neutrinos with ordinary matter, and enable a search for exotic neutrino 
properties. A complete knowledge of the role of 
neutrinos in the Universe in which we live requires a detailed knowledge 
of neutrino masses, mixing, and interactions. 

Our present knowledge of the neutrino and anti-neutrino scattering 
cross sections in matter is limited. The next generation of approved 
neutrino scattering experiments, including MINER$\nu$A \cite{debbie}
in the NuMI beamline and MiniBooNE \cite{Church:1997ry} using neutrinos from the Fermilab Booster,
 are expected to greatly improve our knowledge. 
In particular, within the next few years we anticipate that precise 
measurements will be made of neutrino scattering on nuclear targets. 
However, we will still lack precise measurements of: 
\begin{itemize}
  \item Anti-Neutrino scattering on nuclear targets.
  \item Neutrino and anti-neutrino scattering on nucleon (hydrogen and deuterium) targets.
  \item Neutrino-electron scattering.
\end{itemize}
The anti-neutrino rates per primary proton on target are, depending on energy, a factor 
of 3-5 less than the neutrino rates. The interaction rates on nucleon 
targets are an order of magnitude less than the corresponding rates on 
nuclear targets, and the cross-section for neutrino-electron scattering is considerably smaller than that on nucleons. Hence, beyond the presently approved program, a factor 
of 10-100 increase in data rates will be required to complete the 
neutrino and anti-neutrino scattering measurements. 

Completing the program of neutrino and anti-neutrino scattering measurements is important for two reasons that will be expanded upon in the following sections.

\begin{enumerate}

\item Neutrino scattering experiments can improve our knowledge of the fundamental properties of neutrinos.

\item Neutrino and antineutrino scattering provide a unique tool for probing the structure of matter and obtaining a more complete understanding of the nucleon in general, and its flavor and spin content in particular.

\end{enumerate}

The neutrino scattering program is of interest to a broad community of particle physicists, nuclear physicists, and nuclear-astrophysicists.

\subsection{Fundamental Neutrino Properties}

Neutrino scattering experiments serve to further our understanding of fundamental neutrino properties. Measurements of neutrino-nucleus cross-sections are needed to constrain the systematic uncertainties of neutrino oscillation experiments. Measurements of neutrino-electron scattering, a process with a robust theoretical cross-section, can probe non-standard neutrino properties and couplings.

\subsubsection{Cross-Section Measurements for Oscillation Experiments}

To measure neutrino oscillation probabilities it is necessary to know the 
composition, intensity, divergence,  and energy spectrum of the initial beam, 
and also know all the relevant neutrino and anti-neutrino cross-sections. 
A well designed ``near detector'' setup enables the initial beam to be well 
characterized, and provides cross-section measurements with adequate  
precision for the oscillation program. In practice it has proved necessary 
to have more than one type of near detector. The K2K Experiment used both 
a near detector that replicated the far detector, and additional near detectors 
optimized to learn more about the underlying cross-sections. The Fermilab long 
baseline neutrino program is following a similar strategy. The MINOS near detector replicates the far detector, and the MINER$\nu$A detector \cite{debbie} has been designed to learn 
more about the underlying cross-sections and nuclear effects. 

Neutrino scattering experiments will play a key role in allowing future precision
oscillation experiments to reach their ultimate sensitivity.  To obtain the most precise value of $\Delta m^2_{32}$ (which is
eventually required to extract mixing angles and the CP-violating phase) we must
better understand and quantify the nuclear processes interposed between
the interaction of an incoming neutrino and measurement of outgoing particles in the detector.   Extracting mixing parameters
such as $\theta_{13}$, and ultimately the neutrino mass hierarchy and
CP phase, also requires much better understanding of the neutral current resonant and coherent cross-sections that contribute to the background. 
The precision measurement of nuclear effects and exclusive cross-sections will provide the necessary foundation for the study of neutrino oscillation with high-luminosity beams at the Fermilab Proton Driver. The unprecedented statistical power will otherwise be compromised by systematic uncertainties from poorly known cross-sections.

The same careful study of cross-sections and nuclear effects must be performed with anti-neutrinos to understand matter effects and CP violation.  To approach the same statistical accuracy with anti-neutrinos as with neutrinos, and thus have the same impact on oscillation measurement systematics many more protons on target are needed.  This is due to a factor of 1.5 - 2.0 in the number of $\pi^+$ to $\pi^-$ produced and an additional factor of 2.0 to 3.0 in the cross-section ratio. The MINER$\nu$A proposal assumes about $9 \times 10^{20}$ POT in a 3 ton fiducial volume for neutrino studies.  This implies that for those measurements that are statistics limited, one would need $(30 - 50) \times 10^{20}$ POT for an anti-neutrino scattering experiment to approach similar statistical accuracy.  The combination of a Fermilab Proton Driver and somewhat larger fiducial volume would make this a feasible experiment.

\subsubsection{Neutrino-Electron Scattering}

\begin{figure}
\centering
\includegraphics[width=3.2in]{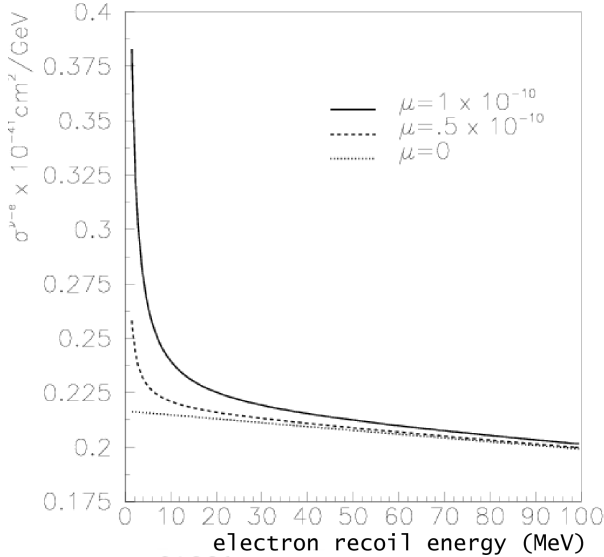}
\caption{Differential cross section versus electron recoil kinetic
energy, T, for $\nu e \rightarrow \nu e $ events.  The electroweak
contribution is linear in T (bottommost line), while
contributions from nonzero neutrino magnetic moments yield
sharp rises at low T. The magnetic moment $\mu$ is given in units of the Bohr magneton $\mu_B$.}
\label{fig:nmm}
\end{figure}
The SM predictions for neutrino-electron elastic scattering  
have little theoretical uncertainty, and a measurement of $\nu e \rightarrow \nu e$ scattering 
can therefore be used to search for physics beyond the SM.  
Since it is now known that neutrinos have non-zero mass, a neutrino magnetic moment becomes a possibility.
Within the SM, modified to include finite neutrino masses, neutrinos 
may acquire a magnetic moment via radiative corrections. With $m_\nu =1$~eV, the 
resulting magnetic moment would be $\sim 3 \times 10^{-19} \mu_B$ where 
$\mu_B = e/2m_e$ is the Bohr magneton.
This value is too small to be detected.  
Hence, a search for a neutrino magnetic moment is a search for physics beyond 
the SM.

The current best limit on the muon neutrino magnetic moment is 
$\mu_{\nu_\mu} \leq 6.8 \times 10^{-10} \mu_B$ from LSND $\nu_\mu e$ elastic 
scattering~\cite{LSNDnmm}.
This sensitivity may be substantially improved by precisely measuring the elastic scattering rate as a function of electron recoil 
energy.  An electromagnetic contribution to the cross section from the magnetic moment 
will show up as an increase in event rate at low electron recoil energies 
(see Fig.~\ref{fig:nmm}). A high statistics measurement, made possible by the Fermilab Proton Driver, would enable a gain in precision of 10-30 over the LSND measurement.
This sensitivity is sufficient to begin to test some beyond-the-
Standard-Model predictions (which can be as large as $10^{-11} \mu_B$ ). The lower neutrino energy means that a beam created by 8 GeV protons (rather than 120 GeV) would be preferred.

\subsection{Fundamental Properties of Matter}

Neutrinos and anti-neutrinos have only weak interactions making them unique probes of the properties of matter at both the nucleon and nuclear level. With the high statistics available from the Fermilab Proton Driver a variety of high precision measurements become possible.

\subsubsection*{Parton Distribution Functions}
     The study of the partonic structure of the nucleon, using the
neutrino's weak probe, will complement the on-going study of this subject
with electromagnetic probes at Jlab.  The unique ability of the neutrino to
``taste'' only particular flavors of quarks enhances any study of parton
distribution functions. With the high statistics and carefully controlled 
beam systematics from the Fermilab Proton Driver, it should
be possible to isolate the contribution of individual quark flavors to the scattering process.

\begin{figure}[hbt]
\centerline{\includegraphics[height=70mm]{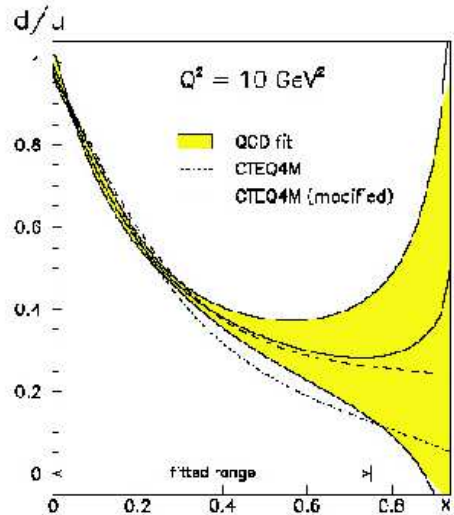}}
\caption{The $d/u$ ratio showing the uncertainty due to nuclear effects in the deuteron.
  Figure taken from Ref.~\cite{Botje}.}
\label{fig:duratio}
\end{figure}

Our knowledge of the parton distributions is still very incomplete. Even the 
valence PDFs are not well known at large $x$.  The ratio of $d/u$ is normally 
determined by comparing scattering from 
the proton and neutron. Since no free neutron target exists, the deuteron 
is used as a neutron target.  This makes little difference at small $x$, but
uncertainties in the nuclear corrections become substantial for $x$ larger
than about 0.6, and make determination of the ratio essentially impossible for $x$ 
larger than about 0.8, as shown in Fig.~\ref{fig:duratio}.
The ratio of $d/u$ can be determined without any nuclear structure effect
corrections by using neutrino and anti-neutrino scattering on hydrogen. Only a MW-scale
Proton Driver can produce anti-neutrino fluxes sufficiently intense for this 
measurement to be made.

\subsubsection*{Generalized Parton Distribution Functions}
One of the most exciting developments in the theory of the structure
of the nucleon has been the introduction of generalized parton 
distributions (GPDs)~\cite{Ji97,Ji97b,Rady96,Rady96b,Collins}.
The usual PDFs are sensitive only to the longitudinal momentum distributions
of the parton.  The GPDs give a more complete picture of the 
nucleon in which the spatial distribution can be determined as a function of 
the longitudinal momentum distribution. However, the GPDs are difficult to 
access experimentally as they require measurement of exclusive final states.
The most promising reaction to date is deeply virtual Compton scattering (DVCS),
 i.e. $p(e,e'\gamma p)$. These measurements are either underway or planned at JLab.
However, a complete determination of the GPDs requires flavor separation 
which can only be accomplished using neutrinos and anti-neutrinos.   
Although MINER$\nu$A will measure GPDs on carbon, nuclear effects will ensure that
this will be considerably inferior to a measurement on the proton.  A true GPD 
measurement would require $p (\bar{\nu}, \mu \gamma n)$ and $n ({\nu}, \mu \gamma p)$ 
reactions using hydrogen and deuteron targets.  Estimates for this
``weak DVCS'' process are currently being made by A. Psaker \cite{Psaker}.  
The CC cross section at 2 GeV is of order $10^{-41}$cm$^2$ and the NC about 10 times
smaller.   These small cross sections will clearly require the higher intensity
neutrino beams that the Fermilab Proton Driver could deliver.

\subsubsection*{Strange Quarks and the Spin Structure of the Nucleon}
The NC elastic scattering of neutrinos and anti-neutrinos
on nucleons ($\nu N \rightarrow \nu N$, $\bar{\nu} N \rightarrow \bar{\nu} N$)
provides information about the spin structure of the nucleon.
In particular, these scattering processes are sensitive to the 
isoscalar spin structure that results from strange quark contributions.
Determination of the strange
quark contribution to the nucleon $\Delta s$ has been a major
component of the JLab program~\cite{Filippone}, but such measurements are
strongly influenced by theoretical assumptions. A precise measurement of NC elastic scattering
would provide a direct measurement of $\Delta s$ that is insensitive to
theoretical assumptions \cite{Bass02}. 
The ideal measurement would be of the 
($\nu p \rightarrow \nu p$)/($\nu n \rightarrow \mu p$) cross-section ratio 
on a deuterium target. 
The ultimate goal in this program requires measuring NC elastic scattering
with both neutrinos and anti-neutrinos, with a large event sample, on {\em nucleon}
targets, which will require a megawatt-scale proton source to produce 
a narrow band neutrino beam of sufficiently high intensity. Note also 
that this measurement should be done using the lower energy neutrino beam from 8 GeV protons (rather than 120 GeV). These lower energies minimize the background of neutrino induced neutrons from the surrounding environment as well as feed down backgrounds from other neutrino interactions.

\subsubsection*{Elastic Form Factors}
The distributions of charge and magnetism within the nucleus can be 
parameterized using two elastic form factors: the 
electric form factor $G_E$ and the magnetic form factor $G_M$. 
For many years it was assumed that both the charge and magnetic 
distributions fall exponentially. However, precision 
measurements at JLab \cite{Jones00,Gayou02} have shown that this is not the case, with the charge appearing to have a broader spatial distribution than that
of magnetism.  Although the reason for this is not understood, 
it does appear to be an indication of angular momentum between the quarks. 
To understand this more deeply it is 
desirable to precisely measure the weak form factor. 
This can be done through parity violation in
electron-nucleon scattering, with limited precision. In neutrino-nucleon elastic
(or quasi-elastic) scattering nearly half the cross section is due to the weak
form factor, making it a much better probe.
Proposed measurements of the weak form factor (e.g. the MINER$\nu$A experiment)
use scattering from nucleons in nuclei.  Although the statistical precision will be
reasonably good, there is an uncertainty in both extracting the form factor
from scattering from a bound nucleon due to final state interactions and other
conventional effects, as well as the possibility of modification of the form factor
by the nuclear medium.  Thus,
measurement of the form factor using neutrino scattering on hydrogen and deuterium 
is essential. This will require the intensity available at a megawatt-scale Proton Driver.

\subsubsection*{Duality and Resonance Production}
Although QCD appears to provide a good description of the strong interaction, 
we have a very poor understanding of the transition from 
the domain where quarks and gluons are the
appropriate degrees of freedom to the domain best described using
baryons and mesons. 
In the region of modest $Q^2$ (1-10 GeV$^2$) the scattering of electrons 
on nucleons is dominated by resonance production, and 
can also be described using the same formalism as DIS.
Experiments at JLab \cite{ioana1} have, quite unexpectedly, found
that the $F_2$ structure function measured in the resonance region
closely follows that measured in the DIS region.
The phenomenon, called quark-hadron duality, has also been observed
in other processes, such as $e^{+}e{-}$ annihilation into hadrons.
The origins of duality are not well understood 
\cite{report,sabine,close,rolf1,CM93,Di86,Be88,Be88_b}. It is expected to exist for neutrino scattering, though it may manifest itself quite differently.
Of particular interest would be a measurement of the ratio of neutron to proton
neutrino structure functions at large {\it x}. The next decade of experiments 
should provide some information on the validity of duality
using neutrinos.  However,  high precision measurements using
anti-neutrinos and nucleon (hydrogen and deuterium) 
targets will be required in order to fully explore the origins of duality and hence the high fluxes of the Fermilab Proton Driver will be needed.

\subsubsection*{Strange Particle Production}
Measurements of the production of strange mesons and hyperons in neutrino NC and CC processes (e.g. $\nu_\mu n \rightarrow \mu^- K^+ \Lambda^0$  
and $\nu_\mu p \rightarrow \nu_\mu K^+\Lambda^0$) 
provide input to test the theoretical models of neutral current neutrino induced strangeness
production~\cite{strange,strange_b}.   In addition,  such strangeness production is a significant background in searches for proton decay based on the  SUSY-inspired proton decay mode  $p \rightarrow \nu K^{+}$. The existing experimental data on these channels consists of only a handful of events from bubble chamber experiments.  
There are plans to measure these reactions using  the existing K2K data and the future MINER$\nu$A data.  MINER$\nu$A will collect a large sample ($\approx 10,000$) of fully constrained $\nu_\mu n \rightarrow \mu^- K^+ \Lambda^0$ events. However, the anti-neutrino measurements will require a more intense (megawatt-scale) proton source.

\cleardoublepage

\section{The Broader Proton Driver Physics Program}

\begin{figure}[h!]
     \centering \includegraphics[width=\textwidth]{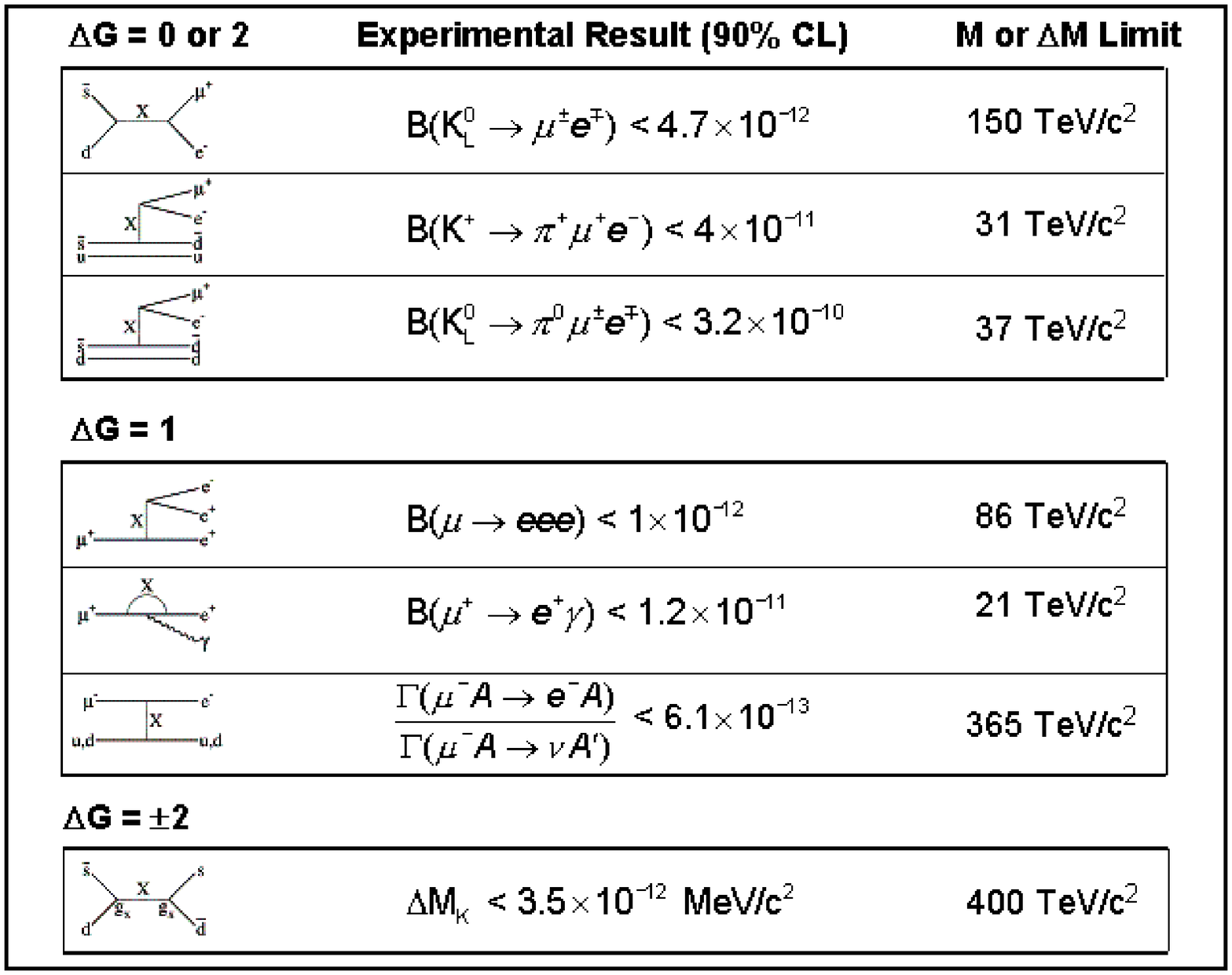}
     \caption {Current limits on Lepton Flavor Violating processes and 
	the mass scales probed by each process.  The upper box is for 
	kaon decays, which involve a change of both quark flavor and 
	lepton flavor.  The bottom box is for muon decays, which involve 
	only lepton flavor change.  The lower limit on the mass scale 
	is calculated assuming the electroweak coupling strength. \label{fig:muonkaon}}

  \end{figure}

  In the past, high precision measurements at low energies have complemented the 
  experimental program at the energy frontier. These low energy experiments not 
  only probe mass scales that are often beyond the reach of colliders, but also 
  provide complementary information at mass scales within reach of the energy 
  frontier experiments. Examples of low energy experiments that have played an 
  important role in this way are muon $(g-2)$ measurements, searches for muon and kaon 
  decays beyond those predicted by the SM, and measurements of rare kaon 
  processes. A summary of the sensitivity achieved by a selection of these 
  experiments is given in Fig.~\ref{fig:muonkaon}. 
    It seems likely that these types of experiment will continue to 
  have a critical role as the energy frontier moves into the LHC era. In particular, 
  if the LHC discovers new physics beyond the SM, the measurement of 
  quantum corrections that manifest themselves in low energy experiments would be 
  expected to help elucidate the nature of the new physics. If no new physics is 
  discovered at the LHC then precision low energy experiments may provide the only 
  practical way of advancing the energy frontier beyond the LHC in the foreseeable 
  future.

  \subsection{Muon Physics}

Solar-, atmospheric-, and reactor-neutrino experiments have established 
Lepton Flavor Violation (LFV) in the neutrino sector, which suggests the 
existence of LFV processes at high mass scales. Depending on its nature, this 
new physics might also produce observable effects in rare muon processes. 
Furthermore, CP violation in the charged lepton sector, revealed for example 
by the observation of a finite muon Electric Dipole Moment (EDM), might be part 
of a broader baryogenesis via leptogenesis picture. Hence, the neutrino 
oscillation discovery enhances the motivation for a continuing program of 
precision muon experiments. 
In addition, the expectation that there is new physics at the TeV scale also 
motivates a new round of precision muon experiments. LFV muon decays and the muon 
anomalous magnetic moment $a_\mu = (g-2)/2$ and EDM are sensitive probes of new 
dynamics at the TeV scale. In general, with sufficient sensitivity, these 
experiments would help elucidate the nature of new physics observed at the LHC. 
As an example, in SUSY models the muon $(g-2)$ 
and EDM are sensitive to the diagonal elements of the slepton mixing matrix, 
while LFV decays are sensitive to the off-diagonal elements. 
If SUSY is observed at the LHC we will probably have some knowledge of the slepton 
mass scale. Precision muon experiments would provide 
one of the cleanest measurements of $\tan \beta$ and of the new CP violating 
phases. It is 
possible that no new physics will be observed at the LHC. In this case 
precision muon experiments might provide, for the foreseeable future, one of 
the few practical ways to probe physics at higher mass scales. Note that the 
Brookhaven $(g-2)$ Collaboration have reported a value for $(g-2)$ that is 2.7 
standard deviations away from the SM prediction. Noting that the 
muon $(g-2)$ is sensitive to any new heavy particles that couple to the muon, it 
is possible that the current measurements are providing an early indication of 
the existence of new TeV-scale particles. Higher precision measurements are 
well motivated.

\begin{table}[t]
\begin{center}
\begin{tabular}{|c|r|r|} \hline
                & \multicolumn{2}{|c|}{Sensitivity}                             \\
Measurement     & Present or Near Future         & Fermilab Proton Driver       \\ \hline
 EDM $d_\mu$    &  $< 3.7 \times 10^{-19}$ e-cm  & $< 10^{-24} - 10^{-26}$ e-cm \\
  $(g-2)$ $\sigma(a_\mu)$ & $0.2 - 0.5$ ppm      &      0.02 ppm                \\
  BR($\mu\rightarrow e\gamma$) &  $\sim 10^{-14}$ &      $\sim 10^{-16}$         \\
  $\mu A \rightarrow e A$ Ratio & $\sim 10^{-17}$         &      $\sim 10^{-19}$         \\ \hline
\end{tabular}
\caption{A comparison of the present or near future sensitivities of the muon experiments considered in the text to those attainable with a Fermilab Proton Driver.}
\label{tab:muon_sens}
\end{center}
\end{table}

In the following, after describing the muon source at the Fermilab Proton Driver, the expected sensitivity of muon EDM, $(g-2)$, and LFV decay experiments is discussed. Table~\ref{tab:muon_sens} summarizes the expected improvements in sensitivity.

\subsubsection{The Muon Source}

\begin{figure}
\begin{picture}(420,65)
\linethickness{2pt}
\put(  0,  0){\includegraphics[width=\textwidth]{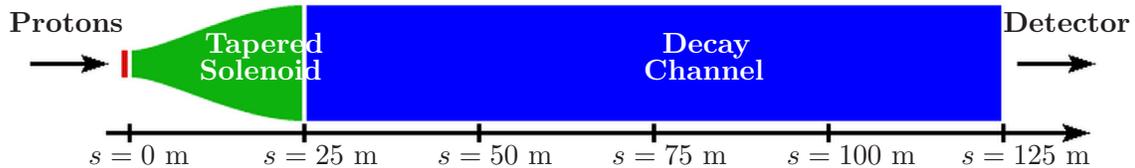}}
\put( 40,  0){\small $s = 0$ m}
\put(106,  0){\small $s = 25$ m}
\put(172,  0){\small $s = 50$ m}
\put(238,  0){\small $s = 75$ m}
\put(304,  0){\small $s = 100$ m}
\put(370,  0){\small $s = 125$ m}
\put( 10, 50){\small \bf Protons}
\put(386, 50){\small \bf Detector}
\put( 84, 42){\small \bf \color{white} Tapered}
\put( 82, 32){\small \bf \color{white} Solenoid}
\put(257, 42){\small \bf \color{white} Decay}
\put(250, 32){\small \bf \color{white} Channel}
\end{picture}
\caption{Schematic of the solenoid-based muon source discussed in the 
text. The performance of this channel has been simulated using the MARS code.}
\label{fig:muonlayout}
\end{figure}

Low energy high precision muon experiments require high intensity beams. 
Since most of the 8~GeV Fermilab Proton Driver beam from the SC linac would not be 
used to fill the MI, it would be available to drive a high intensity muon source. 
In addition to high intensity, precision muon experiments also require an 
appropriate bunch structure, which varies with experiment.
In the post-collider period it might be possible to utilize the Recycler Ring to 
repackage the 8~GeV proton beam, yielding a bunch structure optimized for each 
experiment. The combination of Proton Driver plus Recycler Ring would provide the front-end 
for a unique muon source with intensity and flexibility that exceed any existing 
facility.

The Recycler is an 8 GeV storage ring in the MI tunnel that can run at the same 
time as the MI. The beam from the Fermilab Proton Driver SC linac that is 
not used to fill the MI could be used to fill the Recycler Ring approximately ten 
times per second.  The ring would then be emptied gradually in the 100~ms intervals 
between linac pulses. Extraction could be continuous or in bursts.
For example, the Recycler Ring could be loaded with one linac pulse of 
$1.5 \times 10^{14}$ protons every 100~ms, with one missing pulse every 1.5 seconds 
for the 120 GeV MI program.  This provides $\sim 1.4 \times 10^{22}$ protons at 
8~GeV per operational year 
($10^7$~seconds). In the Recycler each pulse of $1.5 \times 10^{14}$ protons can 
be chopped into 588 bunches of $0.25 \times 10^{12}$ protons/bunch with a pulse 
width of 3~ns.  A fast kicker allows for the extraction of one bunch at a time. 
The beam structure made possible by the Proton Driver linac and the Recycler Ring 
is perfect for $\mu \rightarrow$ e conversion experiments, muon EDM searches and 
other muon experiments where a pulsed beam is required.  Slow extraction from the 
Recycler Ring for $\mu \rightarrow e \gamma$ and $\mu \rightarrow 3e$ searches is 
also possible. 


\begin{figure}[t!]
\begin{center}
\begin{tabular}{|c|c|c|c|c|c|c|c|c|}

\hline

Experiment	&Sensitivity &$\int I_{\mu}dt$  &$I_0/I_m$	
&$\delta$T	&$\Delta$T &
$p_\mu$		& $\Delta p_\mu/p_\mu$ \\

		&Goal	& & & [ns]& [$\mu s$] &[MeV] & [\%] \\
\hline

$\mu A \rightarrow eA$	&$10^{-20}$	&$10^{21}$	&$< 10^{-10}$	&$< 100$
&$> 1$	&$< 80$	&$< 5$ \\

$\mu \rightarrow e\gamma$  &$10^{-16}$     &$10^{17}$      &n/a   &n/a
&n/a  &$< 30$ &$< 10$ \\

$\mu \rightarrow eee$  &$10^{-16}$     &$10^{17}$      &n/a   &n/a
&n/a  &$< 30$ &$< 10$ \\

\hline

$\tau_{\mu}$  &0.5 ppm     &$10^{15}$      &$<10^{-5}$	&100   &$> 20$
& 30 &$< 10$ \\

\hline

$(g-2)$  &0.02 ppm     &$10^{15}$      &$<10^{-7}$ &$< 50$   &$> 10^3$
& 3100 &$< 2$ \\

$EDM$  &$10^{-24}$ e $\cdot$ cm     &$10^{16}$      &$<10^{-6}$ &$< 50$   &$> 10^3$
& $< 1000$ &$< 2$ \\ \hline
\end{tabular}
\includegraphics[width=0.5\textwidth]{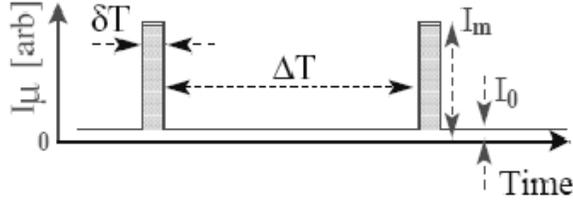}
\end{center}
\caption{
Beam requirements for new muon experiments.  Shown are the expected 
sensitivity goals at the time of the Fermilab Proton Driver, the number of muons 
needed to achieve that sensitivity, the muon suppression between pulses, the
length and separation of pulses and the momentum spread of the muon beam.
}
\label{fig:muon_require}
\end{figure}

The performance of the strawman muon source shown in Fig.~\ref{fig:muonlayout} 
has been simulated using the MARS code. 
The evolution of the pion and muon fluxes down the decay channel is shown in  
Table~\ref{tab:charged}. The scheme will yield $\sim 0.2$ 
muons of each sign per incident 8~GeV proton. With $1.4 \times 10^{22}$ protons at 
8~GeV per operational year (corresponding to $\sim2$ megawatts) this would yield $\sim 3 \times 10^{21}$ muons per year. 
This muon flux greatly 
exceeds the flux required to make progress in a broad range of muon 
experiments (see Fig.~\ref{fig:muon_require}). 
However, the muons at the end of the decay channel have low energy, a large momentum 
spread, and occupy a large transverse phase space. Without further manipulation their 
utilization will be very inefficient. The interface 
between the decay channel and each candidate experiment has yet to be designed. In 
Japan a Phase Rotated Intense Slow Muon Source (PRISM) based on an FFAG ring that 
reduces the muon energy spread (phase rotates) is being designed. This phase 
rotation ring has a very large transverse acceptance ($800\pi$~mm-mrad) and a momentum 
acceptance of $\pm30\%$ centered at 500~MeV/c. PRISM reduces the momentum and momentum 
spread to 68~MeV/c and $\pm 1-2\%$ respectively. Hence, a PRISM-like ring 
downstream of 
the decay channel might accept a significant fraction of the muon spectrum 
and provide a relatively  efficient way to use the available muon flux. Explicit 
design work must be done to verify this, but it should be noted that a muon 
selection system that utilizes only $1\%$ of the muons available at the end of the 
decay channel will still produce an adequate muon flux for most of the cutting-edge 
experiments described in the following sections.

Finally, a new 8~GeV multi-MW Proton Driver at Fermilab, together with an 
appropriate target, pion capture system, decay channel, and phase rotation 
system could provide the first step toward a Neutrino Factory based on a 
muon storage ring. The additional systems needed for a neutrino factory are 
a cooling channel (to produce a cold muon beam occupying a phase space that 
fits within the acceptance of an accelerator) an acceleration system (which 
perhaps would use the Proton Driver SC linac), and a storage ring with long straight 
sections.

\begin{table}[t!]
\begin{center}
\begin{tabular}{|l|c|c|c|c|c||} \hline
         &          &          &          &           &          \\[-0.5cm]
         & $s=25$ m & $s=50$ m & $s=75$ m & $s=100$ m & $s=125$ m\\[0cm]
\hline
         &          &          &          &           &          \\[-0.25cm]
$\mu^{+}/P$ & 0.16     & 0.20     & 0.21     & 0.21      & 0.22     \\[0.25cm]
$\mu^{-}/P$ & 0.16     & 0.20     & 0.21     & 0.21      & 0.21     \\[0.25cm] 
\hline
         &          &          &          &           &          \\[-0.25cm]
$\pi^{+}/P$ & 0.095    & 0.051    & 0.030    & 0.020     & 0.014    \\[0.25cm]
$\pi^{-}/P$ & 0.087    & 0.044    & 0.025    & 0.016     & 0.011    \\[0.25cm] 
\hline \hline
\end{tabular}
\caption{The number of charged particles in the beam per incident 8 GeV primary proton 
as a function of the distance downstream from the target.  These numbers are 
computed using the MARS code. The normalization corresponds to a 2~megawatt Proton Driver.}
\label{tab:charged}
\end{center}
\end{table}

\subsubsection{Electric Dipole Moment}

Electric Dipole Moments violate both  parity (P) 
and time reversal (T) invariance. If CPT conservation is assumed, a finite EDM 
provides unambiguous evidence for CP violation. In the SM EDMs are 
generated only at the multi-loop level, and are predicted to be many orders of 
magnitude below the sensitivity of foreseeable experiments. Observation of a 
finite muon EDM ($d_\mu$) would therefore provide evidence for new CP violating 
physics beyond the SM. CP violation is an essential ingredient of almost all 
attempts to explain the matter-antimatter asymmetry of the Universe. However, 
the measured CP violation in the quark sector is known to be insufficient to 
explain the observed matter-antimatter asymmetry. Searches for new sources of 
CP violation are therefore well motivated.

A number of extensions to the SM predict new sources of CP violation.
Supersymmetric models in which finite neutrino masses and large neutrino mixing 
arise from the seesaw mechanism provide one example. In these models $d_\mu$ can 
be significant. There are examples in which 
the interactions responsible for the masses of right-handed Majorana neutrinos 
produce values of $d_\mu$ as large as $5 \times 10^{-23}$~e-cm. This is well 
below the present limit $d_\mu < 3.7 \times 10^{-19}$~e-cm, but would be within 
reach of a new dedicated experiment at a high intensity muon source. The muon EDM 
group has proposed an experiment and a new beamline at JPARC to obtain a sensitivity of 
$10^{-24}$~e-cm. This sensitivity would still be limited by 
statistics. Higher muon intensities would help, although the measurement would 
also be rate limited. To obtain a sensitivity of O($10^{-26}$)~e-cm would 
require an improved beam structure with many short pulses, each separated by 
at least 500~$\mu$s. Hence, depending on the fate of the JPARC proposal, a 
muon EDM experiment at a Fermilab Proton Driver would be designed to obtain a 
sensitivity somewhere in the $10^{-24}$ - $10^{-26}$~e-cm range.

\subsubsection{Muon $(g-2)$}

The Brookhaven $(g-2)$ Collaboration has reported a value that is 2.7 standard 
deviations from the SM prediction. This could be an early indication 
of new physics at the electroweak symmetry breaking scale. Physics that would affect 
the value of $(g-2)$ include muon substructure, anomalous gauge couplings, 
leptoquarks, and SUSY. For example, in a minimal supersymmetric model with 
degenerate sparticle masses the contribution to  $a_\mu = (g - 2)$/2  
could be substantial, particularly for large $\tan \beta$. 
With degenerate SUSY masses, the estimated range of masses that correspond 
to the observed $a_\mu$ are 100 - 450~GeV for $\tan \beta = 4 - 40$. 
Hence the present $(g-2)$ experiment is 
probing the mass range of interest for electroweak symmetry breaking.
With further data taking the BNL experiment might be 
able to improve sensitivity by a factor of a few. To make progress beyond this 
will require either an upgraded storage ring or a new ring, and a higher 
intensity muon source at, for example, the Fermilab Proton Driver.

The current measurement of $a_\mu$ is accurate to 
0.46 (stat) $\pm$ 0.27 (sys) ppm.  A proposed upgrade to the 
current program could increase the overall precision from 0.5 ppm to 0.2 ppm.  
A new experiment at the Fermilab Proton Driver might improve this 
precision by an order of magnitude, which would be accomplished by reducing the 
systematic error by a factor of seven and increasing the statistical sample 
by a factor of 200. To fully exploit this improvement in experimental precision 
a corresponding improvement in the precision of the theoretical prediction will 
be required, which will need a very high precision measurement of hadronic 
production in $e^+e^-$ collisions from threshold to $\sim 2.5$~GeV.

\subsubsection{Rare Muon Decays}

If a negative muon is stopped in a target it will be captured by an atom and 
then cascade down to the 1$s$ atomic level. In the SM the muon will 
either decay in orbit or will be captured by the nucleus with the emission of 
a neutrino. In LFV models beyond the SM the muon can also 
be converted into an electron in the field of the nucleus 
($\mu \rightarrow e$ conversion) or can undergo non-SM decays 
($\mu \rightarrow e\gamma$, $\mu \rightarrow eee$, ...).
If the SM is extended to include the seesaw mechanism with right-handed 
neutrinos in the mass range $10^{12}$ to $10^{15}$~GeV, the predicted 
LFV decay branching ratios are unobservably small. However in SUSY seesaw models, 
the resulting LFV decay branching ratios can be significant. 
The predicted branching ratios depend upon the origin 
of SUSY breaking. Muon LFV decay searches therefore place constraints 
on SUSY breaking schemes. 

Consider first $\mu \rightarrow e\gamma$. The present limit on the decay 
branching ratio is $1.2 \times 10^{-11}$. Within the context of SUSY models, 
this limit already constrains the viable region of parameter space.  
The MEG experiment at PSI is expected 
to reach a branching ratio sensitivity of $10^{-14}$ by the end of the decade.
The MEG measurement will provide constraints on SUSY parameter space complementary 
to those obtained by the LHC experiments. If MEG observes $\mu \rightarrow e\gamma$ 
then both a higher statistics experiment and new searches for other non-SM decay modes 
will be well motivated. If MEG only obtains a limit, further progress 
will also require a more sensitive experiment. In either case, a 
$\mu \rightarrow e\gamma$ experiment at a new Fermilab Proton Driver would 
provide a way forward provided a high sensitivity experiment can be designed 
to exploit higher stopped muon rates. Progress will depend upon improvements 
in technology to yield improved background rejection and higher rate capability.  
To illustrate the possible gains in the $\mu \rightarrow e\gamma$ 
sensitivity consider an experiment that uses pixel detectors to track both the 
decay positron and the electron-positron pair from the converted photon. 
It has been shown that if BTeV pixel technology is used in an idealized geometry 
with $10\%$ of the coverage of the old MEGA experiment, then a sensitivity comparable 
to the expected MEG sensitivity might be obtained at a beamline of the sort proposed 
by the BNL MECO experiment. 
The main limitation in using BTeV pixel technology would come from scattering 
in the fairly thick detectors. We can anticipate the development of much thinner 
pixel detectors. If the pixel thickness can be reduced by 
a factor of 10, coverage increased 
to $50\%$ of the MEGA coverage, and the readout rate improved by a factor of 
20, then the resulting single event sensitivity would be improved by about two orders 
of magnitude. 

%

Now consider $\mu \rightarrow e$ conversion. 
Within the context of some SUSY extensions to the SM, the 
$\mu \rightarrow e$ conversion rate is related to the $\mu \rightarrow e\gamma$ 
branching ratio:
\begin{equation}
\Gamma (\mu \rightarrow e) \sim  16 \alpha^4 Z^4_{eff} Z |F(q^2)|^2 
BR(\mu \rightarrow e\gamma)
\end{equation}
where $Z$ is the proton number for the target nucleus, $Z_{eff}$ is the effective charge, 
and $F(q^2)$ is the nuclear form factor. The  $\mu \rightarrow e$ conversion rate normalized 
to the muon capture rate in Ti is then given by:
\begin{equation}
R(\mu^- \rightarrow e^- \; Ti) \sim 6 \times 10^{-3} BR(\mu \rightarrow e\gamma)
\end{equation}
The prediction is model dependent, and hence searches for both $\mu \rightarrow e\gamma$ 
and $\mu \rightarrow e$ conversion are well motivated. The apparent suppression of the 
$\mu \rightarrow e$ conversion rate with respect to the $\mu \rightarrow e\gamma$ decay 
rate is in practice compensated by the higher sensitivity achievable for the conversion 
experiments. The PRIME experiment has been 
proposed at JPARC to 
improve the sensitivity to O($10^{-19}$). Proton economics may well determine the fate of 
PRIME. If PRIME is not able to proceed at JPARC it might be accommodated at the Fermilab 
Proton Driver.

%

  \subsection{Other Potential Opportunities}
    A Fermilab Proton Driver that would provide high intensity beams at both 
    MI energies and at 8~GeV would offer tremendous flexibility for the future 
    physics program, and would enable a vigorous experimental endeavor that 
    extends into and beyond the next two decades. 
    In addition to supporting experiments that exploit lepton beams (neutrinos, anti-neutrinos, and muons),
    a Fermilab Proton Driver could support a 
    variety of experiments using secondary hadron beams. 
    Although the possibilities have not all been explored,
    some specific illustrative examples using kaon, pion, neutron, and antiproton 
    beams have been considered.

  \subsubsection{Kaon Experiments using the MI Beam}
Many crucial features of the quark-flavor sector, such as the nature of the couplings, can only be probed indirectly using rare decays.  Examples from the past include the flavor changing neutral current (FCNC) suppression via the GIM mechanism and CP violation, both discovered with K-decays. Of particular importance are the ultra-rare FCNC modes $K^+ \rightarrow \pi^+\nu\nu$ and $K_L \rightarrow \pi^0 \nu\nu$. The SM predictions for these branching ratios are extraordinarily precise, Br$(K^+ \rightarrow \pi^+ \nu\nu) = (8.0 \pm 1.1) * 10^{-11}$ \cite{Buras:2005gr,Isidori:2005xm} and Br$(K_L \rightarrow \pi^0 \nu\nu) = (3.0 \pm 0.6) * 10^{-11}$ \cite{Buchalla:1993bv,Buras:2004uu}. These decays probe quark flavor physics in $s \rightarrow d$ transitions. The $K \rightarrow \pi\nu\nu$ modes are ideally suited for this purpose since they are predicted in the SM with high theoretical accuracy.  The intrinsic theoretical uncertainty on BR($K_L \rightarrow \pi^0 \nu\nu$) is $<$1\% and it is expected to reach 6\% for BR($K^+ \rightarrow \pi^+\nu\nu$), where the charm quark mass uncertainty dominates. The variety of conceivable new physics scenarios involving $K \rightarrow \pi\nu\nu$ is very large.  Within many supersymmetric models, enhancements of between 3 and 10 times larger than SM expectation are possible\cite{Colangelo:1998pm,Buras:1998ed,Buras:1999da}. In generic models of new physics a 10\% measurement of Br$(K_L \rightarrow \pi^0 \nu\nu)$ constrains the new physics scale to exceed 1280 TeV \cite{Bryman:2005xp}.  

The world sample of  $K^+ \rightarrow \pi^+\nu\nu$ consists of 3 candidate events observed by the combined BNL-E787 and BNL-E949 experiments.  The experimental central value for the branching ratio is $1.5^{+1.3}_{-0.9} \times 10^{-10}$, consistent with the SM expectation.  There are currently no observed candidates for $K_L \rightarrow \pi^0 \nu\nu$. A new generation of experiments has been proposed to observe 50-100 of each of these decay modes within the next 10 years. In the $K_L$ sector, the initiatives are KEK391a and the follow-up experiment at JPARC (LOI-05).  In the $K^+$ sector, the initiatives are JPARC-LOI-04,  NA48/3 at CERN, and P940 at Fermilab. In the Proton Driver era, assuming the presently proposed experiments meet their 50-100 event goal, a reasonable next goal would be to carry out measurements at the 1000 event level. The near term experiments are already pushing the limit of detector technology and so progress will require improvements in detection technique. Assuming this is the case the required number of protons on target can be estimated by assuming a KAMI-like beam line and detector for the $K_L \rightarrow \pi^0 \nu\nu$ case, and using efficiency numbers from the KAMI proposal. For the  $K^+ \rightarrow \pi^+\nu\nu$ case, similar quantities can be extrapolated from the P940 proposal.  The required numbers of protons on target are given in Table~\ref{design}. 

\begin{table*}[htb]
\centering
\newcommand{\m}{\hphantom{$-$}}
\newcommand{\cc}[1]{\multicolumn{1}{c}{#1}}
\renewcommand{\arraystretch}{1.2} 
\begin{tabular}{@{}llll}
\hline
Mode  &  Sample Size &  Physics Measurement & POT \\
\hline
$K^{+}$ $ \rightarrow \pi^+\nu\nu$  &  1000     &  3\% of  $|V_{ts}^*V_{td}|$ &  1.5 $\cdot 10^{20} $\\
$K_{L}$ $ \rightarrow \pi^0\nu\nu$  &  1000     &  1.5\% of  $Im (V_{ts}^*V_{td})$ &  1.6 $\cdot 10^{21} $\\
$K_{L}$ $ \rightarrow \pi^0e^+e^-$  &  2 $\cdot 10^{4}$    &  10\% of  $Im(V_{ts}^*V_{td})$ &  2.5 $\cdot 10^{20} $
\\
\hline
\end{tabular}\\[2pt]
\caption{Desired data sample sizes for various kaon physics 
measurements in the Proton Driver era, and the associated 
number of protons on target (POT) needed.\label{design}}
\end{table*}

Recently two additional decay modes have received attention: $K_L \rightarrow \pi^0 ee$ \cite{Buchalla:2003sj} and $K_L \rightarrow \pi^0 \mu \mu$. These decay modes are fully reconstructible, and therefore are significantly easier to identify than $K \rightarrow \pi\nu\nu$. There are no large backgrounds that could ``feed-down'' and fake the signal.  The only serious backgrounds are $K_L \rightarrow ee\gamma\gamma$ and $\mu\mu\gamma\gamma$ which occur with a branching ratio of about $10^{-7}$ and can be reduced by kinematic cuts to obtain an effective residual background level of $\sim 10^{-10}$. Although this exceeds the expected signal by an order of magnitude, the background is flat over the signal region and with sufficient statistics the signal peaks would enable extraction of the branching ratios. 

\subsubsection{Pion Experiments using the 8 GeV Beam}

Pion experiments can provide precision tests of the SM, and help provide a better understanding of the theory of strong interactions. Pion experiments at the Fermilab Proton Driver would use the 8 GeV beam and only need a power of 0.5 megawatts.

Pions are the lightest hadrons. Their decay modes are few and simple, 
and they therefore provide an exquisite
laboratory for testing fundamental symmetries.  
It is generally agreed that the next important step in pion decay physics
is to accurately measure the branching ratio of the decay 
$\pi^+ \rightarrow e^+\nu(\gamma)$ $(\pi_{e2})$ and normalize it to 
$\pi^+ \rightarrow \mu^+\nu(\gamma)$ $(\pi_{\mu 2})$.  The double ratio 
BR($\pi_{e2}$)/BR($\pi_{\mu 2}$) is theoretically clean, 
probes e-$\mu$ universality in weak 
charged decays, and is predicted \cite{Marciano:1993sh} in the Standard
Model to have a value of $(1.2356 \pm 0.0001) \times 10^{-4}$. 
Beyond-the-SM scenarios
typically preserve lepton universality in weak charged decays, and so it
is believed to be a deeply fundamental symmetry.  The current
world average of the double ratio is 
$(1.230 \pm 0.004) \times 10^{-4}$. 
In comparison with lepton universality tests in $\tau$-decays or W-decays, 
the pion system's experimental precision is 3-10 times better and is 
unlikely to be surpassed.

In addition, the $\pi_{e2}$ measurement provides    
the normalization for measurements of 
the decay $\pi^+ \rightarrow e^+\nu_e\pi^0$ decays $(\pi_\beta )$. 
The uncertainty on the  $\pi_{e2}$ measurements dominate the external 
systematic uncertainties on the  $\pi_\beta $ measurement. This is 
of interest because the CKM matrix element $V_{ud}$ can be extracted
cleanly from $\pi_\beta$ measurements. 
The current best $\pi_\beta$ measurement \cite{Pocanic:2003pf} yields $V_{ud} = 0.9728(30)$. 
The world average is $V_{ud} = 0.9738(5)$, which is dominated
by measurements from super-allowed nuclear decays. 
However, in the future, improved 
 $\pi_\beta$ measurements would allow a theoretically cleaner extraction 
of  $V_{ud}$, and improved precision provided the statistical and systematic 
uncertainties can be decreased. Finally, other rare pion decay modes that provide 
opportunities for searches for new physics are 
$\pi^0 \rightarrow 3\gamma$, $\pi^0 \rightarrow 4\gamma$, and $\pi^0 \rightarrow \nu \nu$.

The PIBETA experiment at PSI is the current state-of-the-art charged pion 
decay experiment. PIBETA uses stopped pions.  
Neutral pion decays are studied using the charge-exchange
reaction $\pi^- p \rightarrow \pi^0n$.   
It is believed that the decay-at-rest technique is now 
at its systematic limit. 
At a Proton Driver, progress could be made by using 
decay-in-flight techniques.

Light meson and baryon spectroscopy probes the confinement and
symmetry properties of QCD. Beyond the usual meson and baryon states, QCD
predicts exotic bound configurations of quarks and gluons which include 
hybrids (e.g. $q\overline{q}g$ and $q^3g$), pure gluon states (e.g. $g^2$ and 
$g^3$), and multiquark
states ($q^2{\overline{q}}^2$, $q^4\overline{q}$, ...). Only a small fraction of these exotic states have been observed. Observation of these states, together with measurements of 
their masses and widths, and determination of their 
quantum numbers, is needed to compare to the predictions of lattice gauge
theory, flux tube models, etc. The spectrum of $q \overline{q}$ mesons is 
well known below 1.5~GeV. Above 1.5~GeV the low angular momentum states are 
poorly known. This is the region where many exotic states are expected. 
For example, lattice gauge theory calculations predict glue balls 
with masses from 2 - 5~GeV.

Our knowledge of the baryon spectrum is also  
incomplete. The only reasonably well-known excited light baryon state is the D(1232),
whose properties are known to $\sim 5\%$.  The properties of a few other 
excited states,
the lowest in each partial wave, are known to $30\%$.  Properties of other
`known' states have much larger uncertainties. Therefore higher precision
information is needed and  missing states must be sought in N$\pi\pi$, $\Lambda$K, 
$\Sigma$K scattering experiments.

The nuclear physics community has 
invested heavily in experimental and theoretical programs aimed at better 
understanding QCD. 
The flagship DOE nuclear physics program for spectroscopy currently uses 
the electromagnetic beam facilities at JLab. However, since the production
mechanisms for the various exotic states are not well understood, it
is important to use different types of beam. Indeed, 
the electromagnetic probes available at JLab 
cannot be analyzed without accounting for the hadronic
intermediate states.  Hence, pion beam experiments at an upgraded Proton Driver
would permit progress in light meson and hadron spectroscopy that would complement 
the JLab program. The measurements would also provide tests of lattice QCD, and  
are therefore of interest to the particle physics community. 
The beam and detector requirements for meson and baryon spectroscopy
studies are quite modest.  The low energy secondary pion and kaon beams
derived from the 8~GeV primary proton beam would be used. A high duty cycle
would be desirable, and hence a bunch stretcher would be required.

  \subsubsection{Neutron Experiments using the 8 GeV Beam}

High power proton beams of a few GeV can produce copious numbers of
spallation neutrons. A Fermilab Proton Driver operating at 8~GeV and 2~megawatts 
could produce neutron beams with an intensity that is comparable to those 
from the most intense neutron sources in the world. The Fermilab Proton 
Driver could therefore support one or more specialized neutron experiments 
that, because of their requirements (e.g. pulse structure needs), 
are either unlikely to be supported at existing or planned neutron spallation 
facilities, or could be performed much better at the Fermilab Proton Driver. 
Some examples of candidate experiments that are of particular 
interest to particle physicists are searches for neutron-antineutron 
oscillations, searches for a neutron electric dipole moment (EDM), and 
precision measurements of the neutron lifetime.

Neutron-antineutron oscillations require baryon number violation, with 
a change in baryon number of two units. Searches for neutron-antineutron 
oscillations are therefore complementary to searches for nucleon decay, 
which requires a change in baryon number of one unit. Thus, 
neutron-antineutron oscillation searches provide a unique test of the 
fundamental stability of matter.
The current limit on a possible transition time between 
the free neutron and antineutron is $\sim 10$~years~\cite{Baldo94}. 
This sensitivity is essentially statistics limited.
Since there are no suitable neutron sources available for a new experiment, 
at present there are no concrete plans to improve this sensitivity.
The Fermilab Proton Driver could provide a 
cold neutron source with an average flux equivalent to that of 
a $\sim20$~megawatt research reactor, and enable an estimated 
2-3 orders of magnitude improvement in sensitivity.

Neutron EDM and lifetime experiments can be pursued using ultra cold 
neutrons (UCNs) which have sufficiently low kinetic energy that they can be 
confined in material or magnetic bottles.
However a beam of UCNs does not exist, and the proposed next 
generation experiments must therefore produce UCNs in situ from an incident 
cold neutron beam. 
Recently a novel suggestion for the production of UCNs using a spallation 
target has been undergoing tests at the Los Alamos Neutron Science 
Center (LANSCE) and elsewhere. The new UCN source concept uses 
a small target that is very closely coupled to a solid deuterium UCN 
converter. This makes it possible to significantly increase the density 
of UCN beyond that obtained using traditional cryogenic moderators, and 
would therefore be of benefit to the neutron EDM and lifetime experiments 
which are statistically limited. At the Fermilab Proton Driver one SC linac 
pulse every few seconds could be used to drive a national UCN facility for 
neutron EDM and lifetime experiments, and a variety of other scientific 
programs.

  \subsubsection{Antiproton Experiments using the MI Beam}

The antiproton source at Fermilab is a unique facility. Although 
built and primarily used to collect antiprotons for the Tevatron Collider, 
over the years it has also been used to support a more diverse set of 
experiments which include putting the world's most stringent 
limits on the antiproton lifetime, studying charmonium states, and providing 
the first unambiguous observation of atomic antihydrogen. 
These experiments were performed at the Fermilab Antiproton Accumulator, which 
provides the world's most intense source of antiprotons. 
In the Proton Driver era, beyond the period of Tevatron Collider operations, 
there will exist only one, or possibly two, other antiproton
sources in the world - one at GSI, and possibly one at CERN. The Fermilab antiproton  
source would continue to be the most intense in the world. 
Indeed, the Fermilab Proton Driver would be expected to enable an increase in the 
intensity of the present source by about a factor of two. Hence, the Fermilab 
antiproton source would continue to be a unique facility. The possible antiproton 
experiments that could be pursued in the future have not been exhaustively studied. 
However two examples have been considered: The continuation of quarkonium formation 
studies, and a search for CP violation in hyperon decays.

The study of the charmonium and bottomonium systems has been crucial in
unraveling the short-distance properties of the strong interaction. A 
significant number of important measurements have been made in studies of
antiproton-proton formation of charmonium. Note that a gas-jet target 
experiment in an  antiproton storage ring can 
(i) achieve an energy spread of 10-100 keV  (compared with a few MeV 
in an $e^+e^-$ experiment) allowing precise measurements of heavy-
quarkonium masses and widths, and 
(ii) study the formation of all non-exotic mesons (only $1^{--}$ states 
can be directly formed by $e^+e^-$ annihilation). 
Knowledge of the charmonium and bottomonium spectra is incomplete. 
There are states to discover, and their masses and widths can help us better 
understand the strong interaction. For example, in charmonium the $h_c$ must 
be confirmed, the significant mass discrepancy
between the BELLE and BABAR sightings of the $\eta_c'$ resolved, the
$h_c$ and $\eta_c'$ widths measured, and the other narrow states
identified and characterized, namely the $\eta_{c2} (1^1D_2)$, $\psi_2
(1^3D_2)  [X(3872)?]$, $1^3D_3$, $2^3P_2$, and $1^1F_4$. 

There are only a few particle-antiparticle systems that are experimentally 
accessible and are sensitive to new sources of CP violation. The hyperon-antihyperon 
system, which can be made in a gas-jet target experiment at the Fermilab antiproton 
source, is one example. Seeking a deeper understanding of CP violation is important 
if we are to understand baryogenesis. 
The Standard Model (SM) predicts a slight {\em CP} asymmetry in the decays of
hyperons~\cite{Hyperon-CP_a,Hyperon-CP_b,Hyperon-CP_c,ACP_a,ACP_b,Valencia,Tandean}. 
Physics beyond the SM can result in large enhancements in this CP asymmetry. 
For example, the supersymmetric calculation of He {\it et al.}~\cite{He} 
predicts asymmetries that are up to two orders of magnitude larger than the SM 
prediction. Although as yet unpublished, the Fermilab HyperCP data is expected 
to yield measurements of $\Lambda$ and $\overline{\Lambda}$ decays with sufficient 
sensitivity to observe CP violation in hyperon decays if it is an order of magnitude 
larger than the SM prediction. Provided systematic uncertainties can be controlled, 
an experiment at the Fermilab antiproton source could improve this sensitivity by an 
order of magnitude, and hence be sufficiently sensitive to observe CP violation at 
the level predicted by the SM, and precisely measure any enhancement that might be 
present due to new physics.

\cleardoublepage

\begin{figure}[b!]
\begin{center}
\scalebox{.65}{\rotatebox{270}{\includegraphics*[bb=78 61 532 721,clip]{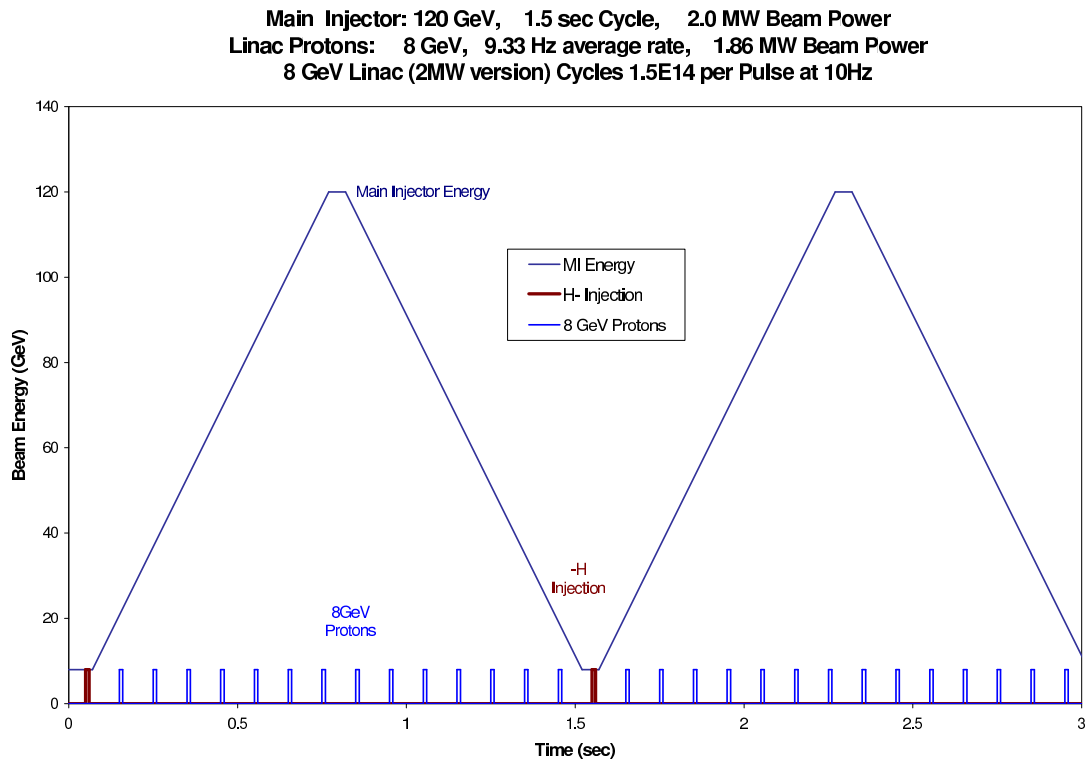}}}
\end{center}
   \caption{\label{fig:MIcycle} Proton Driver bunch structure and the Main Injector cycle.
}
\end{figure}
\section{Compatibilities and Proton Economics}

The proton driver design that is currently favored consists of an 8~GeV $H^-$ 
Linac that initially would produce a 0.5~megawatt beam, and that can eventually be 
upgraded to produce 
a 2~megawatt beam. A small fraction of the 8~GeV beam would be used to fill the MI with the 
maximum beam that, with some modest improvements, it can accelerate. This would 
yield a 2~megawatt beam at MI energies. Hence the upgraded proton source would deliver 
two beams that can be simultaneously used for experiments: a 2~megawatt beam at MI 
energies, and eventually an almost 2~megawatt beam at 8~GeV. To illustrate this the 
cycle structure 
is shown in Fig.~\ref{fig:MIcycle}. The MI would receive one pulse from the Linac 
every 1.5~sec. The cycle time is dominated by the time to ramp up and ramp down 
the MI energy. The 14 Linac pulses that are available, while the MI is ramping 
and at flat top, would provide beam for an 8~GeV program.

The initial NO$\nu$A long-baseline program would be expected to be the primary user 
of the 2~megawatt MI beam. The high-energy neutrino scattering program also needs this 
beam, and would be expected to coexist with NO$\nu$A. The other candidate uses of the 
MI beam include supporting experiments at an antiproton source, and supporting
kaon experiments. The antiproton source could operate in parallel with the 
MI neutrino program with a modest reduction in the available POT for the NuMI beam. 
The kaon program, in contrast to the neutrino program, would require 
a slow extracted beam. However, with an additional storage ring, it is possible 
that a kaon program could be run during neutrino running with only a minor impact 
on the POT available for the neutrino program. This could be accomplished by fast 
extraction of a fraction of the MI bunches and transfer into a stretcher ring. The 
protons stored in the ring would then be slowly extracted for the kaon program. 
A ring in the Tevatron tunnel would be ideal for this purpose.

The candidates for using the 8~GeV beam are a low energy neutrino scattering experiment, 
a pion program, a muon program, and  a neutron program. The neutrino scattering experiment 
requires short pulses with large gaps between pulses so that beam-unrelated backgrounds 
can be suppressed. The pion program requires a beam stretcher to produce long pulses
and hence minimize the instantaneous intensity. Many experiments in the muon program 
require a CW beam with bunches that are short compared to the muon lifetime with gaps 
between bunches of several muon lifetimes. Hence all of these programs will require 
an additional storage ring to manipulate the bunch structure. It is possible that, in 
the post-collider era, this storage ring could be the Recycler. 
Noting that each of these 8~GeV sub-programs requires a different bunch structure
it is natural to consider a scenario in which they run sequentially rather than in 
parallel. To illustrate this, we can imagine that initially an 8~GeV neutrino scattering 
experiment is the primary user of the 8~GeV beam, followed by (or possibly interleaved with) 
one or more pion experiments. In a second phase the facility is 
upgraded to include a low energy muon source, and one or more low energy muon experiments 
become the primary user(s). In a third phase, if required, the muon source could be 
developed to become the front-end of a neutrino factory. This could be a very long-term 
20 - 30 year plan.

\cleardoublepage

\section{Summary}

There is a compelling physics case for a Proton Driver at Fermilab.
This upgrade to the existing accelerator complex is motivated by the
exciting developments in neutrino physics. Independent of the value
of the unknown neutrino mixing parameter $\theta_{13}$, a 2~megawatt
Main Injector proton beam would facilitate, over the coming decades,
one or more long-baseline neutrino experiments that would make critical
contributions to the global neutrino oscillation program. The NuMI beam
is the only neutrino beam in the world with an appropriate energy and
a long enough baseline for matter effects to significantly change the
effective oscillation parameters. With a 2~megawatt Proton Driver this
unique feature of the Fermilab neutrino program can be exploited to:
\begin{itemize}
\item  Probe smaller values of $\theta_{13}$ than can be probed with
reactor-based experiments or with any existing or approved
accelerator-based experiment.
\item If $\theta_{13}$ is not very small, determine the neutrino mass
hierarchy and greatly enhance the sensitivity of the global neutrino
program to CP violation.
\item If $\theta_{13}$ is very small, establish the most stringent limit
on $\theta_{13}$ and prepare the way for a neutrino factory driven by the
Proton Driver.
\end{itemize}
The MI neutrino oscillation physics program would be supplemented with a
broad program of neutrino scattering measurements that are of interest to
particle physicists, nuclear physicists, and nuclear astrophysicists.
The neutrino scattering program would utilize both the Proton Driver
upgraded NuMI beam, and neutrino and anti-neutrino beams generated using
the protons available at 8~GeV. The overall neutrino program motivates the
highest practical primary beam intensities at both MI energies and at 8~GeV.
In practice this means 2~megawatts at MI energies and 0.5-2~megawatts at 8~GeV.

Additional physics programs could also be supported by the Proton Driver. In particular, the Proton Driver could support (i) a program of low energy experiments that probe the TeV mass scale in a way that is complementary to the LHC experiments, and (ii) a program of low energy experiments that are of interest to the nuclear physics community and that is complementary to the JLab program. The possibilities using the 8~GeV beam include (i) the development of a very intense muon source with a bunch structure optimized for (g-2), muon EDM, and LFV muon decay experiments, (ii) a program of low energy pion experiments, and (iii) some specific experiments using spallation neutrons. The possibilities using the MI beam include a program of kaon experiments, and some specific experiments using the antiproton source.

For decades to come, a Fermilab Proton Driver would support an exciting world class neutrino program that would address some of the most fundamental open questions in physics, and could also support a broader program of low energy experiments.

\cleardoublepage

\section{Appendix: Proton Driver Scientific Advisory Group}

 The physics case presented in this document emerged from the "Fermilab Proton Driver Workshop", 6-9 October, 2004, and from subsequent work conducted by the participants. This work, together with the documented physics case, was presented to and reviewed by the Proton Driver Scientific Advisory Group, appointed by the Fermilab Directorate:

\begin{itemize}
\item Peter Meyers, Princeton (Chair)
\item Ed Blucher, Chicago
\item Gerhard Buchalla, Munich
\item John Dainton, Liverpool
\item Yves Declais, Lyon
\item Lance Dixon, SLAC
\item Umberto Dosselli, INFN
\item Don Geesaman, ANL
\item Geoff Greene, ORNL
\item Taka Kondo, KEK
\item Marvin Marshak, Minnesota
\item Bill Molzon, UCI
\item Hitoshi Murayama, UC Berkley
\item James Siegrist, LBNL
\item Anthony Thomas, JLab
\item Taku Yamanaka, Osaka
\end{itemize}

The present document benefited greatly from the comments and suggestions arising from the reviews
conducted by this advisory group.

\cleardoublepage

{\footnotesize
\bibliography{references}
}

\end{document}